\documentclass[%
 reprint,
 superscriptaddress,
 amsmath,amssymb,
 aps,
 prb,
floatfix,
]{revtex4-2}

\usepackage{comment}
\usepackage{graphicx}% Include figure files
\usepackage{dcolumn}% Align table columns on decimal point
\usepackage{newtxtext,newtxmath}
\usepackage{bm}% bold math
\usepackage[hidelinks]{hyperref} % pdfLatex
\hypersetup{% 
 setpagesize=false,
 bookmarksnumbered=true,%
 bookmarksopen=true,%
 colorlinks=true,%
 linkcolor=blue,
 citecolor=blue,
}

\usepackage{mathrsfs}

\def \otoc {\mathcal F}

\begin{document}

\title{Exactly Solvable Disorder-free Quantum Breakdown Model: Spectrum, Thermodynamics, and Dynamics}

\author{Kinya Guan}
\affiliation{Department of Physics, Graduate School of Science, The University of Tokyo, Tokyo~113-0033, Japan}

\author{Hosho Katsura}
\affiliation{Department of Physics, Graduate School of Science, The University of Tokyo, Tokyo~113-0033, Japan}
\affiliation{Institute for Physics of Intelligence, The University of Tokyo, Tokyo 113-0033, Japan}
\affiliation{Trans-Scale Quantum Science Institute, The University of Tokyo, Tokyo 113-0033, Japan}

\date{\today}

\begin{abstract}
We introduce and study a disorder-free version of the quantum breakdown model with all-to-all interactions. The Hamiltonian factorizes into the product of the zero-momentum-mode occupation number and a quadratic Hamiltonian including only pairing terms. This structure makes the model exactly solvable and produces an extensive zero-energy degeneracy.
We obtain the spectrum and partition function in closed form and analyze the thermodynamics, spectral form factor, two-point functions, and a regulated out-of-time-ordered correlator (OTOC). The OTOC is time independent at leading order in the large-$N$ limit, while its time-dependent contributions first appear at order $1/N$. At infinite temperature, the time-dependent correction reduces to a two-point function arising from the zero-momentum component of the fermion operators. The model therefore provides a controlled setting for isolating the consequences of the breakdown interaction in the absence of disorder, spatial structure, and environmental coupling.
\end{abstract}

\maketitle

\section{Introduction}
Nonequilibrium dynamics of quantum many-body systems is one of the central topics in modern condensed-matter physics. A prototypical example is dielectric breakdown in insulators, where a strong electric field drives the system far from equilibrium and induces a sudden onset of conduction. In gases, this process has been studied since the early work of Townsend on ionization avalanches~\cite{townsend1910theory}, and similar avalanche-like breakdown phenomena have been observed in solids, liquids, and chemical reaction networks~\cite{sze1991avalanche,sun2016formation,PRLYamanouchi1999,PRBTaguchi2000,akiyama2000streamer,claycomb2001avalanche}.
From a microscopic viewpoint, however, the breakdown process involves highly excited many-body states and strong interactions, together with disorder and environmental coupling in realistic materials. The enormous Hilbert space and the lack of small parameters make a fully quantum-mechanical description generally difficult.

Most previous studies of dielectric breakdown have therefore been formulated at a phenomenological or semi-classical level using ionization avalanches and discharge plasmas (see, e.g., Refs.~\cite{townsend1910theory,raizer1991gas,lieberman2005principles}). These approaches capture breakdown fields and macroscopic transport properties, but they do not directly offer a fully quantum many-body picture of the breakdown process.

In strongly correlated electron systems, microscopic quantum descriptions of dielectric breakdown have also been developed, particularly in the context of Mott insulators~\cite{fukui1998breakdown,oka2010dielectric,oka2008nonequilibrium}. More recently, Lian introduced a \emph{quantum breakdown model} (QBM) on a one-dimensional chain of spinless fermions with a spatially asymmetric interaction intended to mimic the effect of a strong electric field~\cite{lian2023quantum,chen2024quantum,hu2024quantum,liu2025two,hu2025quantum}. In this model, the interplay between the breakdown interaction, local potentials, and disorder gives rise to a crossover from many-body-localized behavior to quantum-chaotic dynamics, accompanied by Hilbert-space fragmentation and many-body scar states. The QBM therefore provides a concrete setting in which nonequilibrium phenomena associated with dielectric breakdown can be studied within a fully interacting quantum many-body framework.

At the same time, the original QBM incorporates several competing ingredients: the breakdown interactions, a chemical potential, quenched disorder, and explicit spatial structure along the chain. While this richness makes the model relevant to realistic solid-state situations, it also makes it difficult to identify the specific role played by the breakdown interaction itself. In addition, although certain special Krylov subspaces are under analytical control, many quantitative properties of the QBM still rely on exact diagonalization of finite chains. It is therefore natural to seek a minimal analytically tractable model that isolates the breakdown interaction from these additional ingredients.

In this work, we take a step in this direction by constructing an exactly solvable, disorder-free all-to-all version of the quantum breakdown model. Its uniform quartic interaction is reminiscent of the disorder-free Sachdev-Ye-Kitaev (SYK) models with uniform couplings~\cite{ozaki2025disorder}. In the present model, however, the Hamiltonian factorizes into the product of the zero-momentum-mode occupation number and a quadratic pairing Hamiltonian. This factorization allows us to obtain the full energy spectrum and the partition function in closed form and to analyze thermodynamic and real-time correlation functions in a controlled way.

In the SYK model with random couplings, the level statistics and spectral form factor exhibit random-matrix behavior, while the OTOC shows exponential growth associated with scrambling~\cite{sachdev1992gapless,kitaev2015simple,maldacena2016remarks,maldacena2016bound,cotler2017black}. Disorder-free SYK models with uniform couplings are instead integrable and do not exhibit random-matrix behavior in their level statistics or spectral form factors, yet their OTOCs remain time dependent in the large-$N$ limit and exhibit a short exponential-like growth regime~\cite{ozaki2025disorder}.

Despite the formal resemblance to these disorder-free SYK models, the regulated OTOC of our model is time independent at leading order in the large-$N$ limit and does not exhibit such an exponential-like growth regime. Its time-dependent contributions first appear at order $1/N$, and the contribution arising from the zero-momentum component of the fermion operators reduces to a two-point function.

The paper is organized as follows. In Sec.~\ref{sec:model}, we define the disorder-free all-to-all quantum breakdown model and set notation. In Sec.~\ref{sec:exact_solution}, we derive the exact solution by reducing the Hamiltonian to a quadratic pairing problem. In Sec.~\ref{sec:spectrum_static}, we discuss spectral and static properties, including the exact structure of the spectrum, thermodynamics, and the spectral form factor. In Sec.~\ref{sec:dynamics}, we study real-time dynamics, two-point functions, and the regulated out-of-time-ordered correlator.

\section{Model}
\label{sec:model}

\begin{figure}[h]
  \centering
  \includegraphics[width=0.6\linewidth]{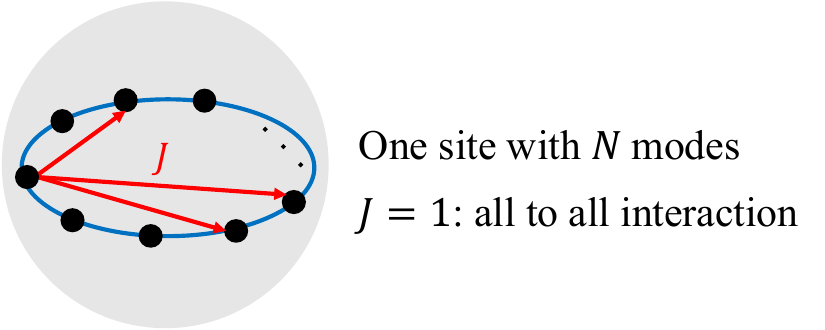}
  \caption{Schematic illustration of the disorder-free all-to-all quantum breakdown model. A single site contains $N$ fermionic modes, and the interaction couples them uniformly with strength $J=1$.}
  \label{fig:alltoall_schematic}
\end{figure}

We consider $N$ spinless complex fermionic modes $c_i,c_i^\dagger$ ($i=0,\ldots,N-1$) obeying
\begin{equation}
\{c_i,c_j^\dagger\}=\delta_{ij},\qquad \{c_i,c_j\}=\{c_i^\dagger,c_j^\dagger\}=0.
\end{equation}
The disorder-free all-to-all quantum breakdown model is described by the Hamiltonian
\begin{equation}
\label{eq:H}
H=\frac{J}{N}\sum_{i=0}^{N-1}\sum_{0\le j<k<l\le N-1}
\Bigl( c_i^\dagger\,c_j c_k c_l \;+\; c_l^\dagger c_k^\dagger c_j^\dagger\, c_i \Bigr),
\end{equation}
where $J$ sets the interaction scale. In the following, we take $J=1$. The overall factor $1/N$ is introduced for convenience. A schematic illustration of the model is shown in Fig.~\ref{fig:alltoall_schematic}.

Let
\begin{equation}
N_f=\sum_{i=0}^{N-1}c_i^\dagger c_i
\end{equation}
denote the total fermion number operator. Since the interaction changes particle number by $\pm2$, the $U(1)$ symmetry is broken, while fermion parity is conserved:
\begin{equation}
[H,(-1)^{N_f}]=0.
\end{equation}

Although the model has no spatial structure, it is useful to arrange the mode labels on an auxiliary ring and introduce the corresponding discrete Fourier modes.

\begin{equation}
f_0=\frac{1}{\sqrt{N}}\sum_{i=0}^{N-1}c_i
\end{equation}
will play a crucial role in the following analysis.
For convenience, we also define the cubic operator
\begin{equation}
Q=\frac{1}{\sqrt{N}}\sum_{0\le j<k<l\le N-1}c_j c_k c_l.
\end{equation}
The Hamiltonian can then be written as
\begin{equation}
\label{eq:Hf0Q}
H=f_0^\dagger Q+Q^\dagger f_0 .
\end{equation}

\section{Exact solution}
\label{sec:exact_solution}
The exact solution follows from factorization
\begin{equation}
\label{eq:Q}
Q=f_0 A,
\end{equation}
where the quadratic fermionic operator $A$ is defined as
\begin{equation}
A=\frac12\sum_{j,k=0}^{N-1}\mathscr A_{jk}\,c_j c_k,
\quad
\mathscr A_{jk}=
\begin{cases}
+1,& j<k,\\
0,& j=k,\\
-1,& j>k.
\end{cases}
\end{equation}
A derivation of Eq.~\eqref{eq:Q} is given in Appendix~\ref{app:exact-solution:Q-factor}.
Substituting Eq.~\eqref{eq:Q} into Eq.~\eqref{eq:Hf0Q}, we obtain
\begin{equation}
\label{eq:HBD1}
H=f_0^\dagger f_0\,A+A^\dagger f_0^\dagger f_0 .
\end{equation}
Let $n_0\equiv f_0^\dagger f_0$.
Since $[n_0,A]\neq0$, Eq.~\eqref{eq:HBD1} cannot yet be written as $H=(A+A^\dagger)n_0$. To identify the terms that survive in Eq.~\eqref{eq:HBD1}, we express $A$ in the Fourier basis.

The full discrete Fourier transform is defined by
\begin{equation}
\label{eq:fq}
f_p=\frac{1}{\sqrt N}\sum_{j=0}^{N-1}c_j e^{-{\rm i}\theta_p j},\quad
\theta_p=\frac{2\pi p}{N}
\end{equation}
where $p=0,1,\ldots,N-1$. These modes satisfy
\begin{equation}
\{ f_p, f^\dagger_{p'} \}=\delta_{p,p'},\qquad \{f_p,f_{p'}\}=0 .
\end{equation}
In this basis, the complete Fourier expansion of $A$ is
\begin{equation}
\label{eq:A}
\begin{aligned}
&A
=
\frac12\sum_{p,q=0}^{N-1}
\tilde{\mathscr A}_{pq}\,f_pf_q,
\\
&\tilde{\mathscr A}_{pq}
=
\begin{cases}
{\rm i}\cot\left({\theta_q}/{2}\right)
\delta_{p+q,\,0\;(\mathrm{mod}\,N)},
& p,q\neq0,\\
1+{\rm i}\cot\left({\theta_p}/{2}\right),
& p\neq0,\ q=0,\\
-1-{\rm i}\cot\left({\theta_q}/{2}\right),
& p=0,\ q\neq0,\\
0,
& p=q=0.
\end{cases}
\end{aligned}
\end{equation}
The Fourier-space kernel is derived in Appendix~\ref{app:exact-solution:Fourier-A}.
Here the key point is that terms containing $f_0$ drop out of~\eqref{eq:HBD1} because
\begin{equation}
n_0 f_0 = f_0^\dagger n_0=0 .
\end{equation}
We therefore define $\tilde A$ by removing the terms with $p=0$ or $q=0$,
\begin{equation}
\label{eq:Aprime}
\tilde A\equiv \frac12\sum_{p,q=1}^{N-1}\tilde{\mathscr A}_{pq}\,f_p f_q ,
\end{equation}
which satisfies $[n_0,\tilde A]=0$. Details are given in Appendix~\ref{app:exact-solution:Aprime}.
As a result,
\begin{equation}
\label{eq:Hre}
H= H_{\rm pair}\,n_0,
\end{equation}
where
\begin{equation}
\label{eq:Hpair}
H_{\rm pair}=\tilde A+\tilde A^\dagger
=\frac{\rm i}{2}\sum_{q=1}^{N-1}\cot\!\left(\frac{\theta_q}{2}\right)
\left(f_{N-q}f_q 
-f_q^\dagger f_{N-q}^\dagger \right).
\end{equation}

The pairing Hamiltonian $H_{\rm pair}$ can be diagonalized by a Bogoliubov transformation (Appendix~\ref{app:exact-solution:Bogoliubov}).
For each independent pair $(q,N-q)$ with $q=1,\ldots,\lfloor (N-1)/2\rfloor$, we introduce the Bogoliubov fermions
\begin{equation}
\gamma_{q\sigma}
=\frac{1}{\sqrt2}\left(f_q-{\rm i}\sigma f_{N-q}^\dagger\right),
\qquad \sigma=\pm1,
\label{eq:bogoliubov}
\end{equation}
which satisfy
\begin{equation}
\{\gamma_{q\sigma},\gamma_{p\tau}^\dagger\}=\delta_{qp}\delta_{\sigma\tau}.
\end{equation}
In terms of these operators,
\begin{equation}
H_{\rm pair}
=\sum_{q=1}^{\lfloor (N-1)/2\rfloor}\sum_{\sigma=\pm1}
\sigma\,\epsilon_q\,n_{q\sigma},
\label{eq:Hfree}
\end{equation}
where $n_{q\sigma}\equiv\gamma_{q\sigma}^\dagger\gamma_{q\sigma}$ and $\epsilon_q\equiv \cot(\theta_q/2)$. Here, $\lfloor\cdot\rfloor$ denotes the floor function. For even $N$, there is also an unpaired mode at $q=N/2$. Since $\epsilon_{N/2}=0$, its occupation does not change the energy; see Appendix~\ref{app:exact-solution:evenN}.

Combining~\eqref{eq:Hre} and~\eqref{eq:Hfree}, we obtain the diagonal form of the full Hamiltonian,
\begin{equation}
\label{eq:Htotal}
H=\left(\sum_{q=1}^{\lfloor (N-1)/2\rfloor}\sum_{\sigma=\pm1}\sigma\,\epsilon_q\,n_{q\sigma}\right)n_0.
\end{equation}
Since $n_0$ has eigenvalues $0$ and $1$, the Hilbert space splits into two sectors. In what
follows, we refer to the $n_0=0$ sector as the \emph{frozen sector} and the
$n_0=1$ sector as the \emph{active sector}.

\section{Spectrum and static properties}
\label{sec:spectrum_static}

Using the factorized Hamiltonian in Eq.~\eqref{eq:Htotal}, we now analyze the energy spectrum, the spectral form factor, and the thermodynamic properties of the model.

\subsection{Spectral structure}
\label{sec:spectrum_structure}

A key feature of the energy spectrum is the extensive zero-energy
degeneracy. Equation~\eqref{eq:Htotal} shows that the spectrum separates according to the eigenvalue of $n_0$.

Let $D_0(N)$ denote the number of zero-energy states for a given system size $N$.
In the frozen sector ($n_0=0$), the energy vanishes identically, so this sector contributes $2^{N-1}$ zero-energy states. In the active sector ($n_0=1$), the many-body energies are
\begin{equation}
E=\sum_{q=1}^{\lfloor (N-1)/2\rfloor}\sum_{\sigma=\pm1}
\sigma\,\epsilon_q\,n_{q\sigma}.
\label{eq:active-spectrum}
\end{equation}
Additional zero-energy states arise when $n_{q+}=n_{q-}$ for all $q$. Combining these with the frozen-sector contribution, we obtain the lower bound on $D_0(N)$:
\begin{equation}
{\underline D}_0(N)=
\begin{cases}
2^{N-1}+2^{(N-1)/2}, & N\ \text{odd},\\
2^{N-1}+2^{N/2}, & N\ \text{even}.
\end{cases}
\label{eq:D0}
\end{equation}
For even $N$, the second term includes an extra factor of $2$ coming from the unpaired zero-energy mode at $q=N/2$.

\begin{figure}[t]
  \centering
  \includegraphics[width=0.90\linewidth]{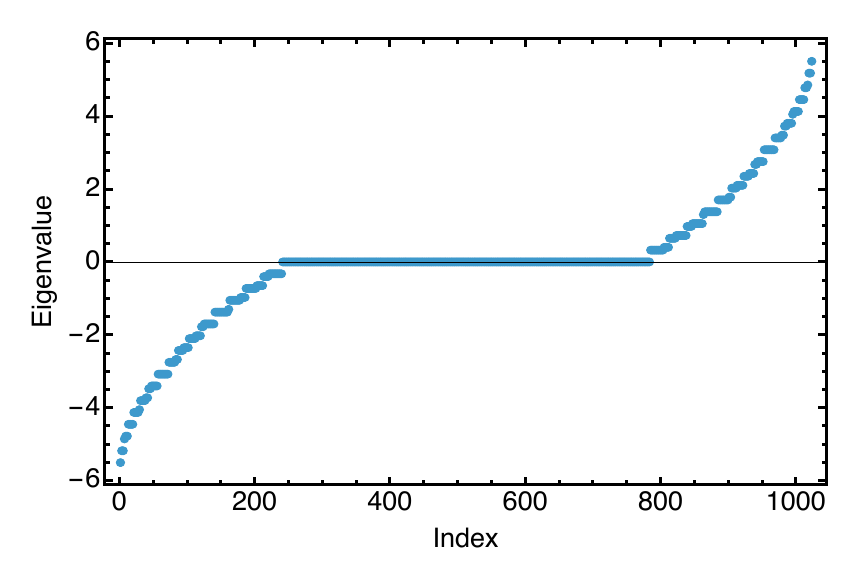}
  \caption{Spectrum for $N=10$. The zero-energy states receive contributions from both the frozen and active sectors, while the nonzero spectrum arises from the active sector.}
  \label{fig:spectrum}
\end{figure}

Figure~\ref{fig:spectrum} shows a representative spectrum (here $N=10$), illustrating the extensive zero-energy degeneracy and the nonzero spectrum arising from the active sector.
Table~\ref{tab:zero-degeneracy} compares the numerical values of $D_0 (N)$ with the lower bound in Eq.~\eqref{eq:D0}. The bound is saturated for all $N$ except for $N=14$. We note in passing that a similar extensive zero-energy degeneracy has been reported in the disorder-free complex SYK model~\cite{iyoda2018}.
\begin{table}[t]
\centering
\begin{ruledtabular}
\begin{tabular}{c c c c c c c c c c c c}
$N$ & $4$ & $5$ & $6$ & $7$ & $8$ & $9$ & $10$ & $11$ & $12$ & $13$ & $14$ \\
\hline
%Exact
$D_0 (N)$ & $12$ & $20$ & $40$ & $72$ & $144$ & $272$ & $544$ & $1056$ & $2112$ & $4160$ & $8324$ \\
%Lower bound
${\underline D}_0 (N)$ & $12$ & $20$ & $40$ & $72$ & $144$ & $272$ & $544$ & $1056$ & $2112$ & $4160$ & $8320$ \\
\end{tabular}
\end{ruledtabular}
\caption{
Exact numbers of zero-energy states, $D_0 (N)$, and the corresponding lower bounds, ${\underline D}_0 (N)$, for different system sizes $N$. 
The discrepancy at $N=14$ arises from an accidental finite-size cancellation among the values of
$\cot(\pi q/N)$, which generates additional zero eigenvalues not accounted for by the generic formula.}
\label{tab:zero-degeneracy}
\end{table}

Finally, we determine the bandwidth $W$ of the nonzero spectrum in the active sector. From Eq.~\eqref{eq:active-spectrum}, with $n_{q\sigma}\in\{0,1\}$, the maximum energy is obtained by setting $n_{q,+}=1$ and $n_{q,-}=0$ for all $q$, while the minimum is obtained by the opposite choice. Therefore,
\begin{equation}
E_{\max}=\sum_{q=1}^{\lfloor (N-1)/2\rfloor}\epsilon_q,
\qquad
E_{\min}=-E_{\max},
\end{equation}
so that
\begin{equation}
W\equiv E_{\max}-E_{\min}
=2\sum_{q=1}^{\lfloor (N-1)/2\rfloor}\cot\!\left(\frac{\pi q}{N}\right).
\end{equation}
For even $N$, the unpaired mode at $q=N/2$ has zero energy and does not affect
the bandwidth. The large-$N$ asymptotics of the cotangent sum are given in
Appendix~\ref{appsec:asymptotics}, which yields $W=O(N\log N)$.

\subsection{Spectral form factor}
\label{sec:sff}

We characterize spectral correlations using the spectral form factor (SFF), defined by
\begin{equation}
g(t,\beta)=\left|\frac{\mathrm{Tr}\,e^{({\rm i}t-\beta)H}}{\mathrm{Tr}\,e^{-\beta H}}\right|^2
=\left|\frac{Z(\beta-{\rm i}t)}{Z(\beta)}\right|^2,
\label{eq:SFF_def}
\end{equation}
where 
\begin{equation}
\label{eq:partfunc}
    Z(\beta)\equiv \mathrm{Tr}\,e^{-\beta H}
\end{equation}
is the partition function. 
In systems with random-matrix-like level correlations, the SFF typically exhibits dip--ramp--plateau behavior~\cite{cotler2017black}.

\begin{figure}[t]
  \centering
  \includegraphics[width=0.9\linewidth]{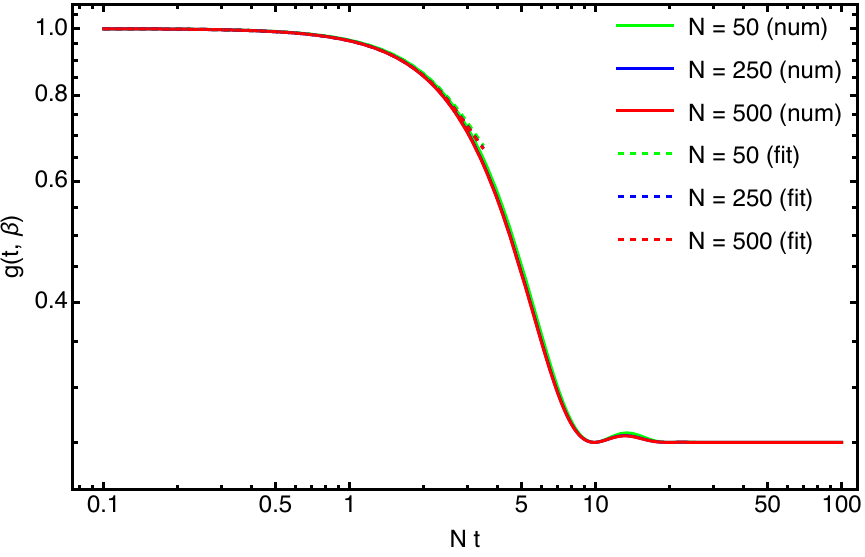}
  \caption{Infinite-temperature SFF $g(t,0)$ for several system sizes $N$. Solid lines show the numerical results, while dashed lines show the small-$t$ expansion in Eq.~\eqref{eq:g_small_t_main}.}
  \label{fig:SFFinfty}
\end{figure}

In the present model, $Z(\beta)$ is available in closed form, so the SFF can be evaluated exactly. 
We first consider the infinite-temperature case $\beta=0$. Figure~\ref{fig:SFFinfty} shows $g(t,0)$ for representative system sizes. The solid lines are numerical results, while the dashed lines show the small-$t$ expansion
\begin{equation}
g(t,0)
=
1
-\frac{t^2}{4N}\binom{N}{3}
+\frac{t^4}{24}\,\frac{1}{4N}
\left[
\frac{7}{2N}\binom{N}{3}^2
+
2\binom{N}{5}
\right]
+O(t^6),
\label{eq:g_small_t_main}
\end{equation}
which captures the initial decay. After the initial decay, the SFF rapidly reaches values close to $1/4$ without developing a clear linear ramp. The frozen sector contributes a strictly time-independent term to $Z(-{\rm i}t)$, whereas the active-sector contribution is suppressed by dephasing. More precisely, at fixed $t>0$, $Z(-{\rm i}t)/Z(0)\to1/2$ as $N\to\infty$, and hence $g(t,0)\to1/4$.

\begin{figure}[t]
  \centering
  \includegraphics[width=\linewidth]{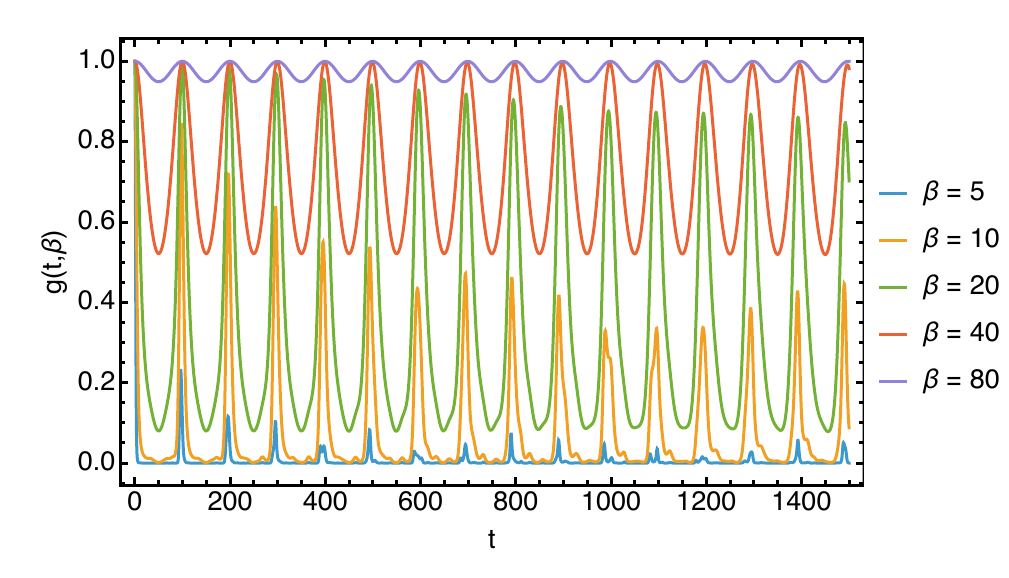}
  \caption{Finite-temperature SFF $g(t,\beta)$ for $N=50$ at several $\beta$. The behavior shows oscillatory revivals rather than a clear ramp, and the revival peaks can vary irregularly in height.}
  \label{fig:SFFfinite}
\end{figure}

At finite temperature, the SFF shows oscillatory revivals rather than a clear ramp, as shown in Fig.~\ref{fig:SFFfinite}. This is qualitatively similar to the behavior found in other integrable models, such as the clean transverse-field Ising chain~\cite{nivedita2020spectral}. In the present model, however, the revival peaks vary irregularly in height, as seen at $\beta=10$.

The derivation of Eq.~\eqref{eq:g_small_t_main} is given in Appendix~\ref{appsec:sff_smallt}, and the large-$N$ analysis of the approach to the plateau is presented in Appendices~\ref{appsec:sff_plateau} and \ref{appsec:PN}.

\subsection{Partition function and thermodynamics}
\label{sec:thermo}
We now derive the partition function, which has already appeared in the SFF through Eq.~\eqref{eq:SFF_def}.
As discussed earlier, 
in the frozen ($n_0=0$) sector, the Hamiltonian vanishes, so the contribution to $Z(\beta)$ is purely entropic.
In the active ($n_0=1$) sector, the Hamiltonian is a sum of independent modes with energies $\pm\epsilon_q$.
For odd $N$, the partition function can therefore be written in closed form as
\begin{align}
Z(\beta)
&= 4^{\lfloor (N-1)/2\rfloor}
+\prod_{q=1}^{\lfloor (N-1)/2\rfloor}\Bigl(1+e^{+\beta\epsilon_q}\Bigr)\Bigl(1+e^{-\beta\epsilon_q}\Bigr)\nonumber\\
&=4^{\lfloor (N-1)/2\rfloor}
+\prod_{q=1}^{\lfloor (N-1)/2\rfloor}4\cosh^2\!\left(\frac{\beta\epsilon_q}{2}\right).
\label{eq:Zbeta}
\end{align}
For even $N$, the unpaired $q=N/2$ mode contributes an overall factor of $2$ 
regardless of $\beta$, which we absorb into the normalization constant since it cancels out in normalized quantities such as the SFF. For simplicity, we restrict 
our analysis below to odd $N$.

From $Z(\beta)$, we obtain the free energy $F(T)=-\beta^{-1}\log Z(\beta)$ and the entropy $S(T)=-\partial F/\partial T$, where $T=1/\beta$. 
For finite systems, the $N$-dependence of these quantities will be made explicit
when needed.

\begin{figure}[t]
  \centering
  \includegraphics[width=0.9\linewidth]{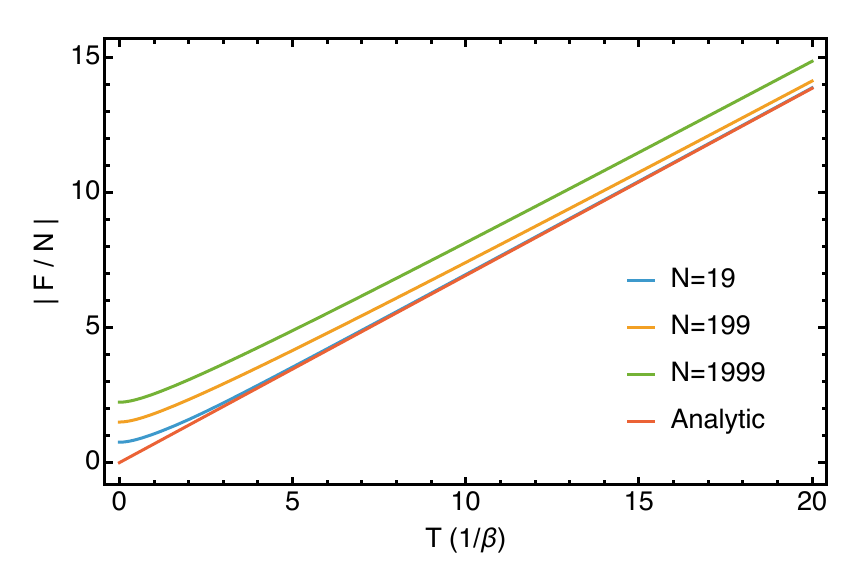}
  \caption{Free energy per mode $|F(T,N)|/N$ as a function of $T=1/\beta$ in the high-temperature regime.
For each fixed $N=19,199,1999$, the numerical data approach the asymptotic line $T\log 2$ as $T$ increases.}
  \label{fig:F_highT}
\end{figure}

Two limits are particularly transparent: the high- and low-temperature limits.
In the high-temperature limit $\beta\to 0$, Eq.~\eqref{eq:Zbeta} gives $Z(\beta)\to 2\cdot 4^{\lfloor (N-1)/2\rfloor}$, hence
\begin{equation}
\frac{F(T,N)}{N}\to -T\log 2,
\qquad
\frac{S(T,N)}{N}\to \log 2,
\label{eq:highT}
\end{equation}
for each fixed $N$ as $T\to\infty$.
Because no Kac-type rescaling is introduced in the present normalization~\cite{Campa2009,Defenu2023}, larger $N$ does not necessarily imply a closer agreement with $T\log 2$ at the same temperature.
Figure~\ref{fig:F_highT} compares the numerical free energy per mode with the asymptotic line $T\log 2$.

At low temperature, we first determine the large-$N$ asymptotic behavior at fixed $\beta$ and then consider $\beta\to\infty$. In this order of limits, the active sector dominates, and the leading behavior is
\begin{equation}
\frac{F(T,N)}{N}
= -\frac{1}{\pi}\bigl(\log N+\gamma-\log\pi\bigr)
-\frac{\pi}{6}T^2+O(T^4),
\label{eq:lowT}
\end{equation}
where $\gamma$ is the Euler--Mascheroni constant. The first term represents the
zero-temperature contribution in this order of limits. Its slow $\log N$
dependence is nonuniversal and originates from the small-$q$ tail
$\epsilon_q=\cot(\pi q/N)\sim N/(\pi q)$, which makes the sum over single-particle
energies scale as $N\log N$ (Appendix~\ref{app:cot-sums}). This logarithmic growth reflects the same normalization choice, namely the absence of Kac-type rescaling~\cite{Campa2009,Defenu2023}.

The corresponding entropy density is
\begin{equation}
  \frac{S(T,N)}{N}=\frac{\pi}{3}T+O(T^3).
\end{equation}

\begin{figure}[t]
  \centering
  \includegraphics[width=0.9\linewidth]{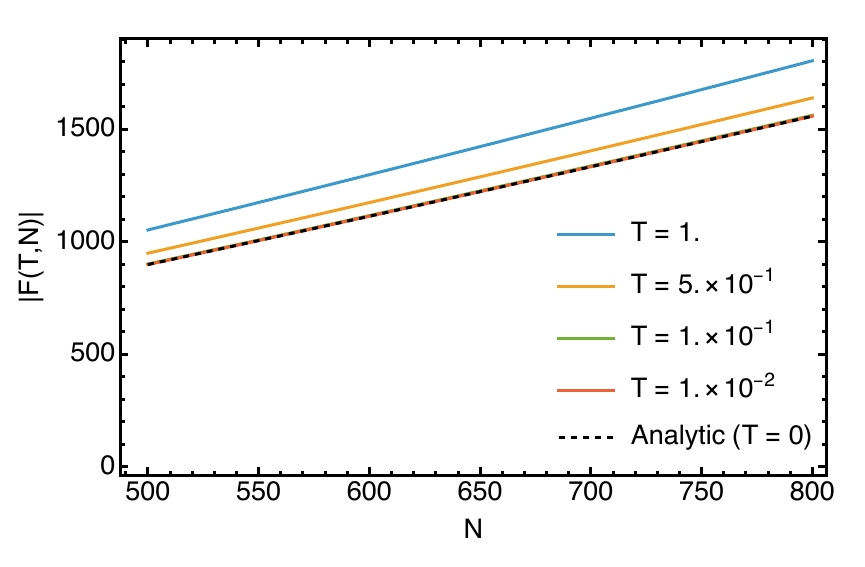}
  \includegraphics[width=0.9\linewidth]{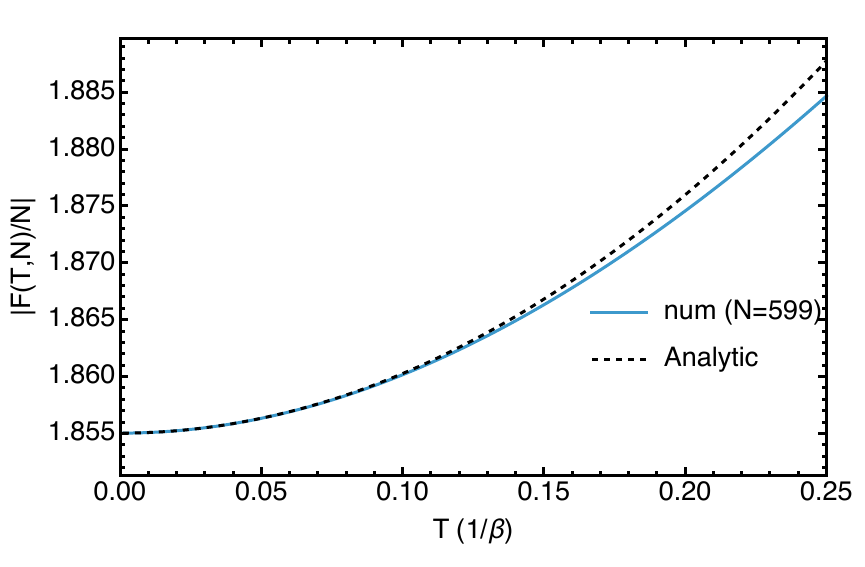}
  \caption{Free energy in the low-temperature regime.
  Upper panel: 
  Free energy $|F(T,N)|$ as a function of $N$ for several temperatures, approaching the asymptotic form
  $\frac{N}{\pi}\bigl(\log N+\gamma-\log\pi\bigr)$ shown by the dashed line.
  Lower panel: 
  Free energy per mode $|F(T,N)|/N$ as a function of $T=1/\beta$ at fixed $N=599$. 
  The solid lines show numerical results, and the dashed line shows the analytical
prediction in Eq.~\eqref{eq:lowT}.}
  \label{fig:F_panels}
\end{figure}

The temperature-dependent correction $-(\pi/6)T^2$ in Eq.~\eqref{eq:lowT} and the linear-in-$T$ entropy are controlled by modes near $\theta=\pi$, where $\epsilon(\theta)=\cot(\theta/2)$ vanishes linearly. Their leading temperature dependence therefore has the same form as that of a $1+1$-dimensional system with a linear low-energy dispersion~\cite{cardy1984conformal,blote1986conformal, affleck1986universal}.
Figure~\ref{fig:F_panels} compares these asymptotic forms with numerical results.
The derivation uses the asymptotics of cotangent sums in Appendix~\ref{appsec:asymptotics}
and the low-temperature expansion given in Appendix~\ref{appsec:lowT_thermo}.

\section{Dynamics and correlations}
\label{sec:dynamics}

The factorized form of the Hamiltonian in Eq.~\eqref{eq:Hre} greatly simplifies the analysis of the real-time dynamics in a particularly transparent way.
We introduce the projectors onto the two sectors distinguished by the eigenvalues of $n_0$:
\begin{equation}
P_0\equiv 1-n_0,\qquad P_1\equiv n_0.
\label{eq:P01_main}
\end{equation}
An operator $\mathcal O$ can be decomposed into the four components $P_\alpha\mathcal O P_\beta$, with $\alpha,\beta=0,1$. The diagonal components preserve the zero-momentum-mode occupation, whereas the off-diagonal components connect the two sectors. For an operator that preserves the zero-momentum-mode occupation, its frozen-sector component is time independent, while its active-sector component evolves under the quadratic Hamiltonian $H_{\rm pair}$. The exact projector representation of the time evolution of a general operator is given in Appendix~\ref{app:U-sector}.

We consider thermal expectation values $\langle\cdots\rangle=\mathrm{Tr}[\rho(\beta) \,\cdots]$, where $\rho(\beta)=e^{-\beta H}/Z(\beta)$ and $Z(\beta)$ is the partition function defined in Eq.~\eqref{eq:partfunc}. The sector decomposition of the density matrix is derived in Appendix~\ref{app:rho-sector}. 
For any operator $\mathcal O$, we define
\begin{equation}
\langle \mathcal O\rangle^\alpha
\equiv
\frac{\mathrm{Tr}\!\left(P_\alpha e^{-\beta H}\,\mathcal O\right)}
{\mathrm{Tr}\!\left(P_\alpha e^{-\beta H}\right)},
\qquad \alpha=0,1,
\label{eq:sector_exp_def}
\end{equation}
together with the corresponding sector weights
\begin{equation}
w_\alpha(\beta,N)\equiv \frac{\mathrm{Tr}\!\left(P_\alpha e^{-\beta H}\right)}{Z(\beta)},
\qquad w_0+w_1=1.
\label{eq:sector_weight_def}
\end{equation}
The full thermal expectation value is then
\begin{equation}
\langle \mathcal O\rangle
=w_0\,\langle \mathcal O\rangle^{0}+w_1\,\langle \mathcal O\rangle^{1}.
\label{eq:sector_mix_main}
\end{equation}
Here the superscript $\alpha$ specifies the projector inserted in the thermal trace.
The second term of Eq.~\eqref{eq:Zbeta}, namely the active-sector contribution to the partition function, will be denoted by $Z_{\rm pair}(\beta)$ and is given by
\begin{equation}
Z_{\rm pair}(\beta)
=
\prod_{q=1}^{\lfloor (N-1)/2\rfloor}
\Bigl(1+e^{+\beta\epsilon_q}\Bigr)\Bigl(1+e^{-\beta\epsilon_q}\Bigr).
\label{eq:Zpair_main}
\end{equation}
With this notation,
\begin{equation}
w_1(\beta,N)=\frac{Z_{\rm pair}(\beta)}{4^{\lfloor (N-1)/2\rfloor}+Z_{\rm pair}(\beta)}.
\label{eq:w1_main}
\end{equation}
In each fixed sector, the density matrix is Gaussian. In the frozen sector, it reduces to an infinite-temperature ensemble of the $(N-1)$ nonzero-momentum modes, 
whereas in the active sector it is determined by the quadratic Hamiltonian
$H_{\rm pair}$. Wick's theorem can therefore be applied to products of nonzero-momentum operators within each fixed sector. Correlators involving components that connect the two sectors require separate treatment. The momentum-space correlators and their extension to complex time are given in Appendix~\ref{app:2pt-momentum}.

\subsection{Two-point functions}
\label{subsec:two_point}

We study finite-temperature two-point functions in real time, $\langle X_j(t)Y_k(0)\rangle$, where $X,Y\in\{c,c^\dagger\}$, together with their counterparts in the Fourier basis $f_p$ defined in Eq.~\eqref{eq:fq}.

The operators $f_q$ with $q\neq0$ preserve the zero-momentum-mode occupation. In the active sector, $H_{\rm pair}$ separates into independent Bogoliubov pairs $(q,N-q)$. For $q\neq 0$ (and $q\neq N/2$ when $N$ is even), one finds
\begin{align}
\big\langle f_q^\dagger(t)\,f_q(0)\big\rangle^{1}
&=\frac12\Bigl[\cos(\epsilon_q t)
-{\rm i} \tanh \Bigl(\frac{\beta\epsilon_q}{2}\Bigr)\sin(\epsilon_q t)\Bigr],
\label{eq:ffnorm}\\
\big\langle f_q(t)\,f_{N-q}(0)\big\rangle^{1}
&=-\frac12\Bigl[\sin(\epsilon_q t)
+{\rm i} \tanh \Bigl(\frac{\beta\epsilon_q}{2}\Bigr)\cos(\epsilon_q t)\Bigr],
\label{eq:ffanom}
\end{align}
with $\epsilon_q=\cot(\theta_q/2)$. The remaining components follow from Hermitian conjugation and fermionic antisymmetry and are given in Appendix~\ref{app:2pt-momentum}.

\begin{figure}[t]
  \centering
  \includegraphics[width=0.9\linewidth]{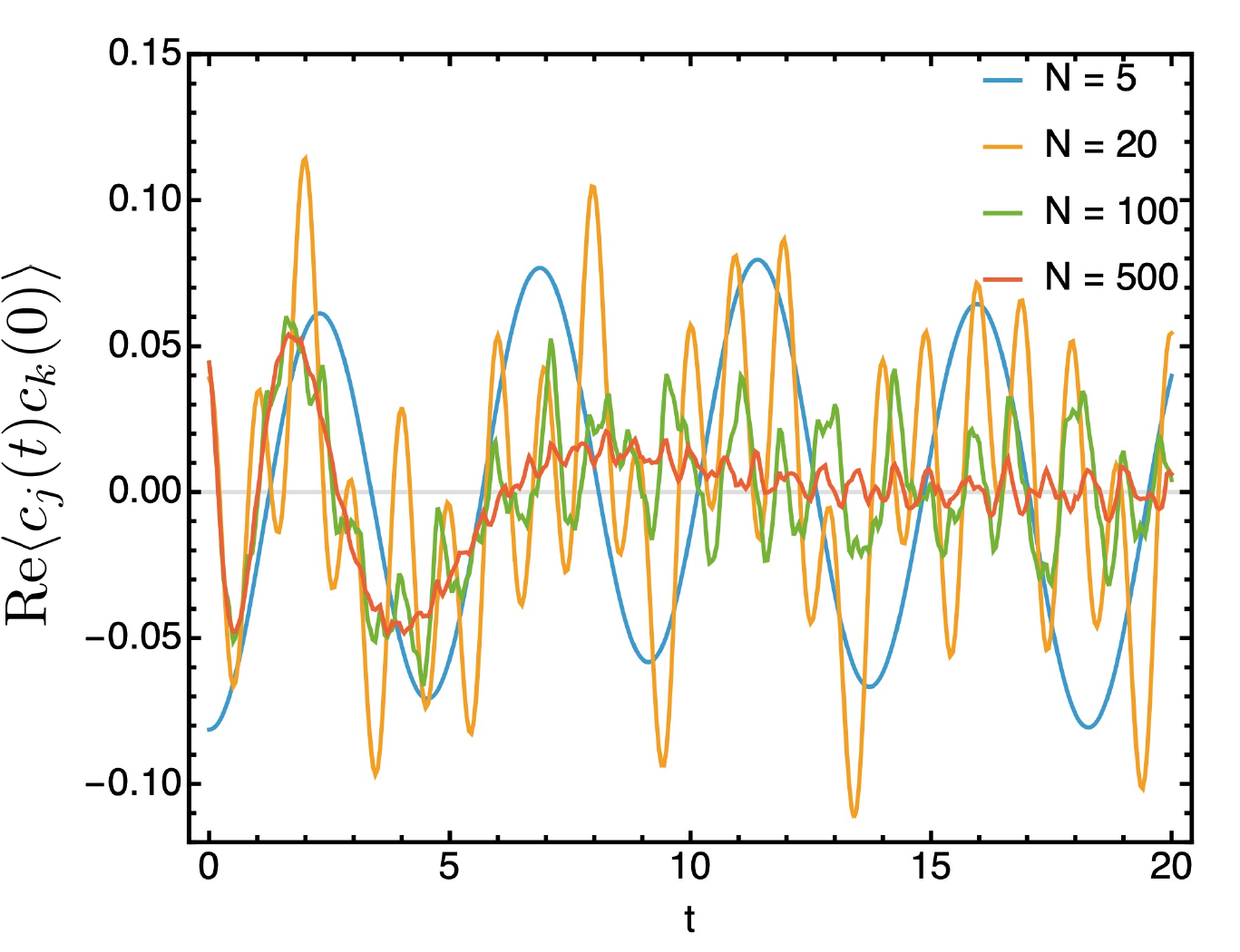}
  \caption{Time evolution of $\mathrm{Re}\langle c_j(t)c_k(0)\rangle$ at separation $j-k=4$ and fixed inverse temperature $\beta=1$, shown for several system sizes $N$.}
  \label{fig:Ccc_diffN}
\end{figure}

Transforming back to the original $c_j$ basis gives compact expressions for the correlators in terms of mode sums.
For $\langle c_j(t)c_k(0)\rangle$, the zero-momentum mode gives no contribution, and we obtain
\begin{align}
\langle c_j(t)\,c_k(0)\rangle
&={\rm i}\,\frac{w_1}{N}\sum_{q=1}^{\lfloor (N-1)/2\rfloor}
\sin\!\big[\theta_q(j-k)\big]\,\nonumber\\
%&\qquad
& \times\Bigl[-\sin(\epsilon_q t)
-{\rm i} \tanh \Bigl(\frac{\beta\epsilon_q}{2}\Bigr)\cos(\epsilon_q t)\Bigr].
\label{eq:Ccc}
\end{align}
The corresponding expressions for
$\langle c_j^\dagger(t)c_k^\dagger(0)\rangle$,
$\langle c_j(t)c_k^\dagger(0)\rangle$, and
$\langle c_j^\dagger(t)c_k(0)\rangle$,
are given in Appendix~\ref{app:2pt-realspace}; see Eqs.~\eqref{eq:Cdagdag_app}, \eqref{eq:CcCdag_app}, and \eqref{eq:CdagC_app}. Equation~\eqref{eq:Ccc} is a weighted superposition over the active-sector frequencies. As $N$ increases, the denser set of frequencies produces the smoother behavior shown in Fig.~\ref{fig:Ccc_diffN}.

In the large-$N$ limit, the mode sum in Eq.~\eqref{eq:Ccc} becomes
\begin{align}
\langle c_j(t)c_k(0)\rangle
&\xrightarrow[N\to\infty]{}
{\rm i}\,\frac{w_1}{2\pi}\int_0^\pi d\theta\,
\sin\!\big[\theta(j-k)\big]\nonumber\\
&\qquad \times
\Bigl[
-\sin\bigl(\epsilon(\theta)t\bigr)
-{\rm i}\tanh\Bigl(\frac{\beta\epsilon(\theta)}{2}\Bigr)
\cos\bigl(\epsilon(\theta)t\bigr)
\Bigr],
\label{eq:Ccc_largeN_main}
\end{align}
where $\epsilon(\theta)=\cot\!\left(\frac{\theta}{2}\right)$. The large-$N$ expressions for the other correlators are given in Appendix~\ref{app:dyn-largeN}. Figure~\ref{fig:CccNinfty} shows
$\mathrm{Re}\langle c_j(t)c_k(0)\rangle$ in the large-$N$ limit at separation $j-k=4$ and several temperatures $T$. The correlator first exhibits oscillatory behavior and then approaches a small-amplitude tail.

\begin{figure}[t]
  \centering
  \includegraphics[width=0.9\linewidth]{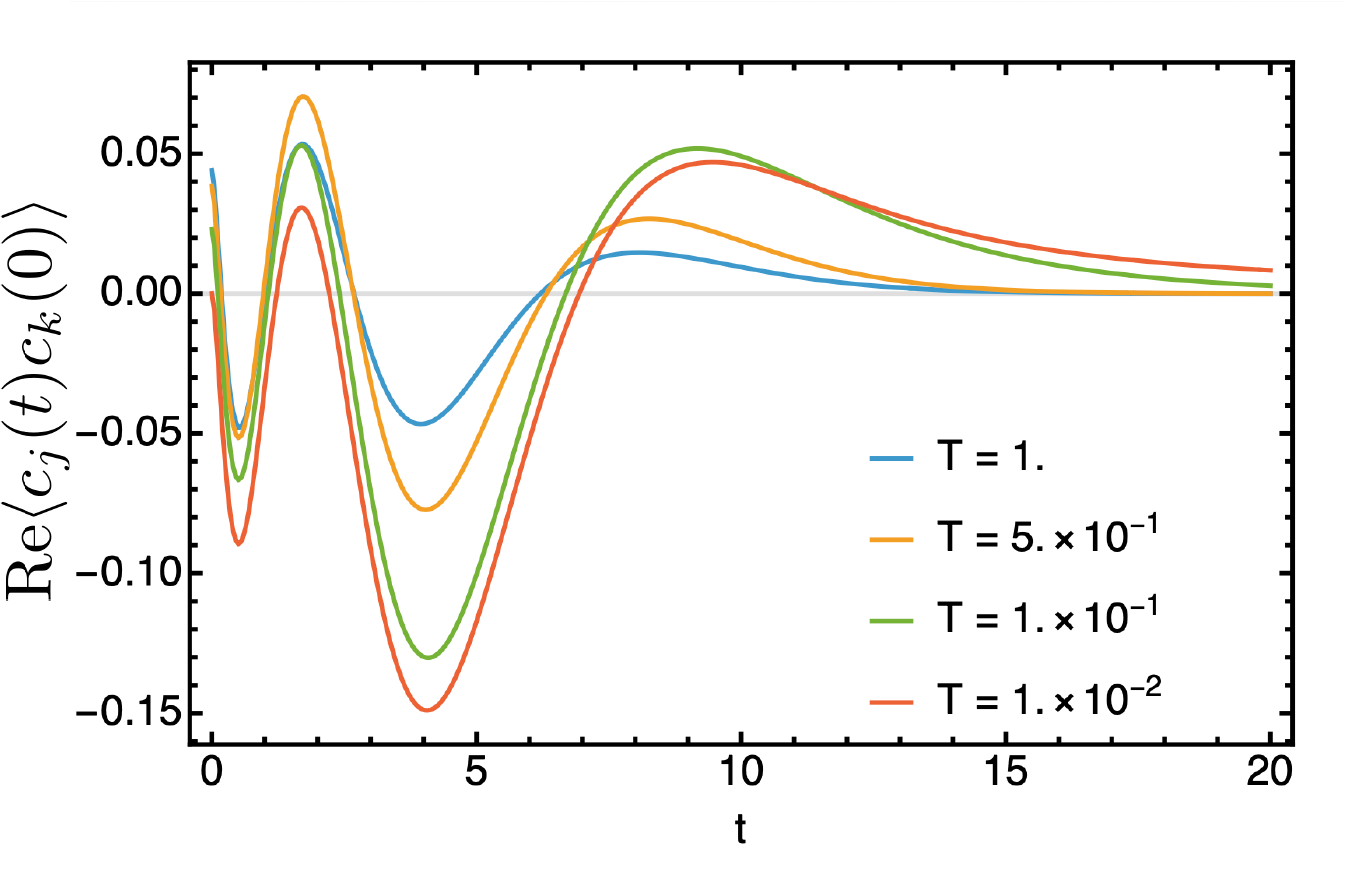}%
  \caption{Time evolution of $\mathrm{Re}\langle c_j(t)c_k(0)\rangle$ in the
  large-$N$ limit for $j=5$ and $k=1$, shown for several temperatures $T$.}
  \label{fig:CccNinfty}
\end{figure}

\subsection{Regulated OTOC}
\label{subsec:otoc}

We study an out-of-time-ordered correlator (OTOC) built from fermionic operators $c_j$, using the standard $\rho(\beta)^{1/4}$ regularization \cite{larkin1969quasiclassical,maldacena2016bound,maldacena2016remarks,hosur2016chaos,afshar2020flat}. We define
\begin{align}
&\otoc(t;\beta)\nonumber\\
&\quad=\frac{1}{N^2}\sum_{j,k=0}^{N-1}
\mathrm{Tr}\!\Bigl[
\rho^{\frac{1}{4}}\,c_j^\dagger(t)\,\rho^{\frac{1}{4}}\,c_k^\dagger(0)\,
\rho^{\frac{1}{4}}\,c_j(t)\,\rho^{\frac{1}{4}}\,c_k(0)
\Bigr].
\label{eq:OTOC}
\end{align}
Using cyclicity of the trace and imaginary-time translation, the regulated OTOC can be written as a thermal four-point function with complex time arguments,
\begin{align}
&\otoc(t_1,t_2,t_3,t_4;\beta)\nonumber\\
&\qquad=\frac{1}{N^2}\sum_{j,k=0}^{N-1}
\mathrm{Tr}\!\Bigl[
\rho\,c_j^\dagger(t_1)\,c_k^\dagger(t_2)\,c_j(t_3)\,c_k(t_4)
\Bigr],
\label{eq:F4pt}
\end{align}
evaluated at
\begin{equation}
(t_1,t_2,t_3,t_4)=\bigl(t-{\rm i}\beta/4,\;0,\;t+{\rm i}\beta/4,\;{\rm i}\beta/2\bigr),
\label{eq:otoc_times}
\end{equation}
where $\otoc(t;\beta)\equiv \otoc(t_1,t_2,t_3,t_4;\beta)$.

The sector decomposition becomes nontrivial for the four-point function in Eq.~\eqref{eq:F4pt}, because each $c_j$ contains both zero- and nonzero-momentum components. We therefore decompose $c_j$ as
\begin{equation}
c_j=d_j+z_0,
\label{eq:c_d_z0}
\end{equation}
where
\begin{equation}
d_j
=
\frac{1}{\sqrt N}
\sum_{q=1}^{N-1}
e^{{\rm i}\theta_qj}f_q,
\qquad
z_0=\frac{1}{\sqrt N}f_0.
\label{eq:d_z0}
\end{equation}
Here $d_j$ contains only nonzero-momentum modes and preserves $n_0$, whereas $z_0$ and $z_0^\dagger$ change it.

Substituting Eq.~\eqref{eq:c_d_z0} into Eq.~\eqref{eq:OTOC}, we write
\begin{equation}
\otoc(t;\beta)
=
\sum_{A,B,C,D\in\{d,z\}}
\otoc_{ABCD}(t;\beta),
\label{eq:OTOC_component_sum}
\end{equation}
where
\begin{align}
&\otoc_{ABCD}(t;\beta)\nonumber\\
&\qquad\equiv
\frac{1}{N^2}
\sum_{j,k=0}^{N-1}
\mathrm{Tr}\!\Bigl[
\rho^{1/4}
A_j^\dagger(t)
\rho^{1/4}
B_k^\dagger(0)
\rho^{1/4}
C_j(t)
\rho^{1/4}
D_k(0)
\Bigr].
\label{eq:OTOC_component_def}
\end{align}
The labels $A,B,C,D$ follow the order of the four operators in Eq.~\eqref{eq:OTOC}. For $\ell\in\{j,k\}$, the label $d$ selects $d_\ell$, whereas $z$ selects $z_0$.

While the sum on the RHS of Eq.~\eqref{eq:OTOC_component_sum} contains $2^4=16$ terms, the algebra of the zero-momentum mode and the sums over $j$ and $k$ leave only three nonvanishing terms:
\begin{equation}
\otoc(t;\beta)
=
\otoc_{dddd}(t;\beta)
+
\otoc_{dzdz}(t;\beta)
+
\otoc_{zdzd}(t;\beta).
\label{eq:OTOC_surviving_terms}
\end{equation}
The other terms vanish due to the properties
$f_0^2=(f_0^\dagger)^2=0$ and
$\sum_{j=0}^{N-1}d_j=0$.

To evaluate the terms in Eq.~\eqref{eq:OTOC_surviving_terms}, we note that $\otoc_{dddd}$ can be separated into frozen- and active-sector contributions, and Wick's theorem can be applied in each sector. The same separation cannot be used for $\otoc_{dzdz}$ and $\otoc_{zdzd}$ because $n_0$ changes between successive operator insertions. These terms are therefore evaluated by inserting $P_0+P_1$ between successive operators. They are identical and reduce to the same two-point function. As a result, we obtain
\begin{align}
&\otoc(t;\beta)
=
w_0\,\otoc_{dddd}^{0}(t;\beta)
+
w_1\,\otoc_{dddd}^{1}(t;\beta)
\nonumber\\
&\quad
-\frac{2Z_{\rm pair}(\beta/2)}
{N\left[4^{\lfloor(N-1)/2\rfloor}+Z_{\rm pair}(\beta)\right]}\nonumber\\
&\qquad\times\left[
\frac{1}{N}
\sum_{q=1}^{\lfloor(N-1)/2\rfloor}
\frac{\cos(\epsilon_qt)}
{\cosh(\beta\epsilon_q/4)}
+
\frac{1+(-1)^N}{2}\,
\frac{1}{2N}
\right].
\label{eq:OTOC_finiteN}
\end{align}
Here $\otoc_{dddd}^{0}$ and $\otoc_{dddd}^{1}$ denote the frozen- and active-sector contributions, respectively. The complete finite-$N$ calculation is given in Appendix~\ref{app:otoc-evaluation}.

Figure~\ref{fig:OTOC_finite_temperature} shows the real part of the finite-$N$ OTOC in Eq.~\eqref{eq:OTOC_finiteN} at several inverse temperatures. At later times, the correlator approaches a temperature-dependent value with small residual oscillations.

\begin{figure}[h]
  \centering
  \includegraphics[width=\linewidth]
  {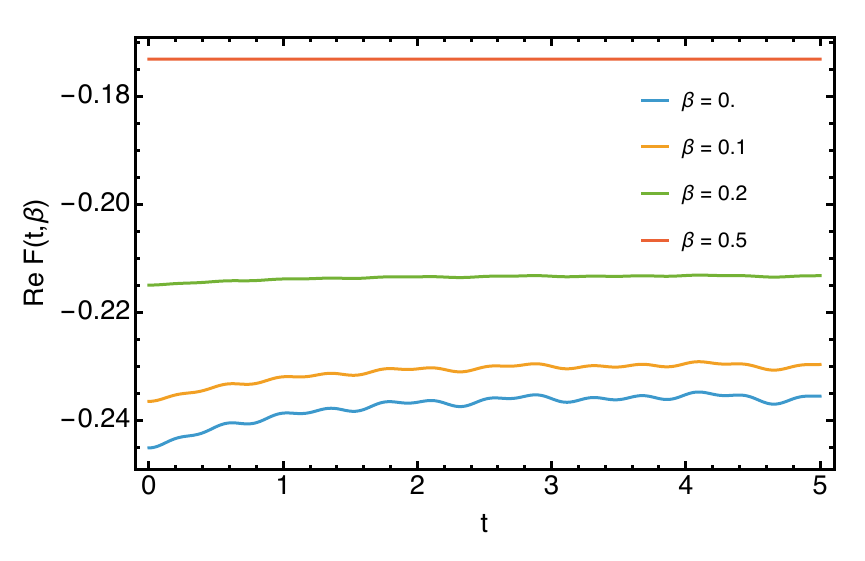}
  \caption{Real part of the regulated OTOC for $N=51$ at several inverse temperatures $\beta$, obtained from Eq.~\eqref{eq:OTOC_finiteN}.}
  \label{fig:OTOC_finite_temperature}
\end{figure}

\begin{figure}[h]
  \centering
  \includegraphics[width=\linewidth]
  {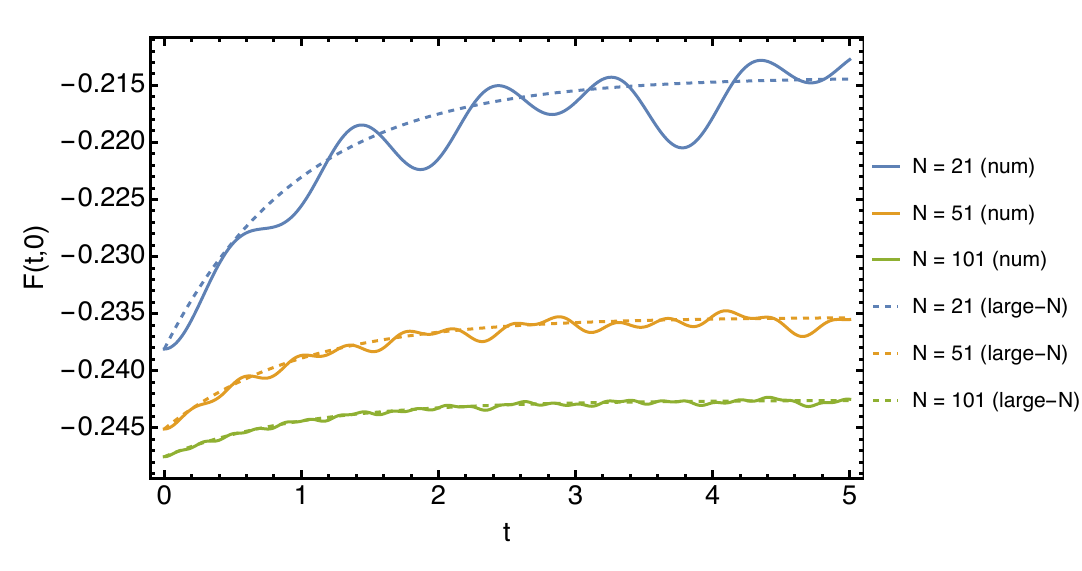}
  \caption{Infinite-temperature regulated OTOC for several system sizes. Solid lines show the numerical evaluation of the exact finite-$N$ expression, while dashed lines show the large-$N$ expansion in Eq.~\eqref{eq:OTOC_beta0_largeN}.}
  \label{fig:OTOC_beta0_largeN}
\end{figure}

Equation~\eqref{eq:OTOC_finiteN} shows that the OTOC has no time-dependent contribution of $O(1)$ in the large-$N$ limit. At infinite temperature, its large-$N$ expansion is
\begin{equation}
\otoc(t;0)
=
-\frac14
+
\frac{1}{N}
\left[
\frac34-\frac12e^{-|t|}
\right]
+
O\!\left(\frac{1}{N^2}\right).
\label{eq:OTOC_beta0_largeN}
\end{equation}
Thus the time dependence first appears at order $1/N$. Figure~\ref{fig:OTOC_beta0_largeN} compares the exact finite-$N$ result with the expansion in Eq.~\eqref{eq:OTOC_beta0_largeN}.
The expansion captures the smooth part of the time dependence, while the exact result retains oscillations associated with the discrete mode frequencies. The agreement improves as $N$ increases.

\section{Conclusion and discussion}
\label{sec:discussion}
We have constructed a disorder-free all-to-all quantum breakdown model that is exactly solvable and admits closed-form expressions for its spectrum, thermodynamics, and real-time correlators. The key structural feature is the factorization of the Hamiltonian through the zero-momentum-mode occupation $n_0=f_0^\dagger f_0$. This factorization is responsible for the extensive zero-energy degeneracy and renders the spectral and dynamical properties analytically tractable.

The comparison with the disorder-free SYK model with uniform couplings provides insight into the short exponential-like OTOC regime found in that model~\cite{ozaki2025disorder}. Under the quartic SYK Hamiltonian, an initially single-fermion operator develops terms in which it is multiplied by occupation operators of other modes. The four-point function therefore contains contributions from many such operator components. In our model, projecting onto the sector with $n_0=1$ leaves the nonzero-momentum fermion operators evolving linearly under $H_{\rm pair}$, without generating analogous many-mode components. The comparison thus suggests that the many-body operator evolution retained by the quartic SYK Hamiltonian is an important ingredient in its short exponential-like OTOC growth and is precisely what the zero-momentum-mode factorization removes in the present model.

It is also useful to compare this sector picture with the original QBM. There, special modes and reference configurations restrict the allowed breakdown processes and thereby lead to fragmentation and constrained breakdown dynamics~\cite{lian2023quantum,chen2024quantum,hu2024quantum,liu2025two,hu2025quantum}. From this viewpoint, the label $n_0$ plays a similar role in the present model. In both cases, simple sector information determines whether a given part of the Hilbert space is frozen or supports nontrivial dynamics. The analogy is not exact, because the present model has no explicit spatial structure, but it still offers a useful perspective on the original QBM, especially beyond analytically tractable subsectors.

This interpretation becomes even clearer in a disorder-free version of the one-dimensional breakdown interaction. For uniform couplings, the same Fourier-transform trick used here isolates a distinguished mode on each site, analogous to $f_0$. This gives a direct interpretation of the resulting Hilbert-space fragmentation; see Appendix~\ref{app:gate-to-1d}. The present all-to-all model can therefore be viewed as a minimal solvable model that isolates the breakdown-type interaction and shows explicitly how sector structure alone can produce markedly different spectral and dynamical behavior.

These observations suggest natural future directions. One direction is to keep the model disorder-free but increase the interaction complexity. For example, several clean breakdown- or SYK-like blocks could be coupled into a chain. This would allow us to test when a more pronounced ramp emerges in the spectral form factor and whether the OTOC exhibits exponential growth. 
Another direction is to study random all-to-all breakdown-type interactions, closer in spirit to the original SYK model. Their spectral and dynamical properties could then be compared with those of the solvable case studied here. Such comparisons may help clarify more generally how breakdown-type interactions manifest themselves in nonequilibrium quantum dynamics.

\begin{acknowledgments}
We would like to thank Kohei Kawabata for drawing our attention to the quantum breakdown model, and Hironobu Yoshida for helpful discussions. 
K.G. was supported by Forefront Physics and Mathematics Program to Drive Transformation (FoPM), a World-leading
Innovative Graduate Study (WINGS) Program, the University of Tokyo. 
H.K. was supported by JSPS KAKENHI Grants No. JP23K25783 and No. JP23K25790, and MEXT KAKENHI Grant-in-Aid for Transformative Research Areas A “Extreme Universe” (KAKENHI Grant No. JP21H05191).

\end{acknowledgments}

\section*{Data Availability}
The code used to perform the calculations and generate the figures shown in this study is openly available on Zenodo \cite{guan_zenodo_qbm_2026}.

\appendix

\section{Technical details of the exact solution}
\label{app:exact-solution}

\subsection{%Factorization 
Derivation of $Q=f_0A$}
\label{app:exact-solution:Q-factor}
Starting from the definitions of $Q$ and $f_0$ in Sec.~\ref{sec:model}, we obtain
\begin{align}
f_0A
&=\frac{1}{\sqrt N}\sum_i c_i\sum_{j<k}c_jc_k \nonumber\\
&=\frac{1}{\sqrt N}
\left(
\sum_{i<j<k}
+\sum_{j<i<k}
+\sum_{j<k<i}
\right)c_ic_jc_k \nonumber\\
&=\frac{1}{\sqrt N}\sum_{0\le i<j<k\le N-1}c_ic_jc_k
=Q,
\end{align}
which proves Eq.~\eqref{eq:Q}.

\subsection{Fourier transform of $\mathscr A_{jk}$}
\label{app:exact-solution:Fourier-A}

Here we derive the Fourier-space kernel $\tilde{\mathscr A}_{pq}$ given in Eq.~\eqref{eq:A}. Define
\begin{equation}
\tilde{\mathscr A}_{pq}
=
\frac{1}{N}\sum_{j,k=0}^{N-1}
\mathscr A_{jk}
e^{{\rm i}\theta_p j}e^{{\rm i}\theta_q k},
\qquad
\theta_p=\frac{2\pi p}{N}.
\end{equation}
Since $\mathscr A_{jk}=-\mathscr A_{kj}$, exchanging $j$ and $k$ gives $\tilde{\mathscr A}_{pq}=-\tilde{\mathscr A}_{qp}$, and hence $\tilde{\mathscr A}_{00}=0$.

Let $z\equiv e^{{\rm i}2\pi/N}$, so that $e^{{\rm i}\theta_pj}=z^{pj}$. For $q\neq0$ and fixed $j$, we have
\begin{align}
\sum_{k=0}^{N-1}\mathscr A_{jk}z^{qk}
=
\sum_{k=j+1}^{N-1}z^{qk}
-\sum_{k=0}^{j-1}z^{qk}
=
\frac{(1+z^q)z^{qj}-2}{1-z^q}.
\label{eq:Ak_sum}
\end{align}
Multiplying by $z^{pj}$ and summing over $j$, we obtain
\begin{align}
\tilde{\mathscr A}_{pq}
&=
\frac{1+z^q}{1-z^q}
\frac{1}{N}\sum_{j=0}^{N-1}z^{(p+q)j}
-
\frac{2}{1-z^q}
\frac{1}{N}\sum_{j=0}^{N-1}z^{pj}
\nonumber\\
&=
\frac{1+z^q}{1-z^q}
\delta_{p+q,\,0\;(\mathrm{mod}\,N)}
-
\frac{2}{1-z^q}\delta_{p,0},
\qquad q\neq0.
\label{eq:Atilde_delta}
\end{align}
Using
\begin{equation}
\frac{1+z^q}{1-z^q}
=
{\rm i}\cot\left(\frac{\theta_q}{2}\right),
\qquad
\frac{2}{1-z^q}
=
1+{\rm i}\cot\left(\frac{\theta_q}{2}\right),
\end{equation}
Eq.~\eqref{eq:Atilde_delta} gives, for $p,q\neq0$,
\begin{equation}
\tilde{\mathscr A}_{pq}
=
{\rm i}\cot\left(\frac{\theta_q}{2}\right)
\delta_{p+q,\,0\;(\mathrm{mod}\,N)}.
\end{equation}
Setting $p=0$ in Eq.~\eqref{eq:Atilde_delta} and using antisymmetry gives
\begin{equation}
\tilde{\mathscr A}_{0q}
=
-\tilde{\mathscr A}_{q0}
=
-1-{\rm i}\cot\left(\frac{\theta_q}{2}\right),
\qquad q\neq0.
\end{equation}
These results establish Eq.~\eqref{eq:A}.

\subsection{Zero-momentum-mode occupation and the commutators}
\label{app:exact-solution:Aprime}

In the main text we replace $A$ by $\tilde A$ defined in Eq.~\eqref{eq:Aprime} by
discarding all contributions with $p=0$ or $q=0$. Here we justify this step.

First, note that the zero-momentum-mode occupation $n_0=f_0^\dagger f_0$ satisfies
\begin{equation}
n_0 f_0 = %0,\qquad 
f_0^\dagger n_0=0,
\end{equation}
so any term in $A$ containing $f_0$ drops out of $n_0A$ and $A^\dagger n_0$ in
Eq.~\eqref{eq:HBD1}. Equivalently, using $[n_0,f_p]=-\delta_{p,0}f_0$ one finds
\begin{align}
[n_0,A]
&=
-\frac12\sum_{p,q}\tilde{\mathscr A}_{pq}
\bigl(\delta_{p,0}f_0f_q+\delta_{q,0}f_pf_0\bigr),
\end{align}
so $[n_0,A]$ arises only from terms containing the zero mode $f_0$.
Dropping all contributions with $p=0$ or $q=0$ therefore yields
\begin{equation}
[n_0,\tilde A]=0,
\end{equation}
and the Hamiltonian becomes
\begin{equation}
H=n_0\tilde A+\tilde A^\dagger n_0 = n_0\,(\tilde A+\tilde A^\dagger)\equiv H_{\rm pair}\,n_0,
\end{equation}
which is Eq.~\eqref{eq:Hre} in the main text.

\subsection{Bogoliubov diagonalization of $H_{\rm pair}$}
\label{app:exact-solution:Bogoliubov}

To diagonalize Eq.~\eqref{eq:Hpair}, we use the fact that $\tilde A$ in Eq.~\eqref{eq:Aprime} couples only pairs of nonzero-momentum modes $(q,N-q)$. For $q\neq0$ and, when $N$ is even, $q\neq N/2$, define $\epsilon_q=\cot(\theta_q/2)$ and write
\begin{align}
H_{\rm pair}
&=\frac{\rm i}{2}\sum_{q=1}^{N-1}\epsilon_q\left(f_{N-q}f_q+f_{N-q}^\dagger f_q^\dagger\right)\nonumber\\
&={\rm i}\sum_{q=1}^{\lfloor (N-1)/2\rfloor}\epsilon_q\left(f_{N-q}f_q-f_q^\dagger f_{N-q}^\dagger\right),
\end{align}
where we paired $q$ with $N-q$ and used $\epsilon_{N-q}=-\epsilon_q$.

We introduce the Nambu spinor $\Psi_q=(f_q,\ f_{N-q}^\dagger)^T$ to express $H_{\rm pair}$ in $2\times2$ matrix form for each $q$:
\begin{equation}
H_{\rm pair}
=\sum_{q=1}^{\lfloor (N-1)/2\rfloor}\Psi_q^\dagger
\begin{pmatrix}
0 & -{\rm i}\epsilon_q\\
{\rm i}\epsilon_q & 0
\end{pmatrix}
\Psi_q.
\end{equation}
This is diagonalized by the Bogoliubov modes
\begin{equation}
\gamma_{q\sigma}=\frac{1}{\sqrt2}\left(f_q-{\rm i}\sigma f_{N-q}^\dagger\right),
\qquad \sigma=\pm1,
\end{equation}
which satisfy $\{\gamma_{q\sigma},\gamma_{p\tau}^\dagger\}=\delta_{qp}\delta_{\sigma\tau}$ and yield
\begin{equation}
H_{\rm pair}=\sum_{q=1}^{\lfloor (N-1)/2\rfloor}\sum_{\sigma=\pm1}\sigma\,\epsilon_q\,n_{q\sigma},
\qquad n_{q\sigma}=\gamma_{q\sigma}^\dagger\gamma_{q\sigma},
\end{equation}
which reproduces Eq.~\eqref{eq:Hfree}.

\subsection{Even-$N$ unpaired mode and Hilbert-space dimension}
\label{app:exact-solution:evenN}

When $N$ is even, the mode $q=N/2$ is unpaired and $\epsilon_{N/2}=\cot(\pi/2)=0$. It contributes only a twofold degeneracy independent of energy. For odd $N$, the active sector contains $(N-1)/2$ independent pairs and thus $N-1$ Bogoliubov modes; together with the contribution of the zero-momentum mode, the total Hilbert-space dimension is $2\times2^{N-1}=2^N$. 
For even $N$, the active sector contains $N/2-1$ independent pairs giving $N-2$ Bogoliubov modes, and the unpaired $q=N/2$ mode contributes an additional factor $2$, so the total dimension is again $2\times2^{N-2}\times2=2^N$.

\section{Cotangent-sum identities and asymptotics}
\label{app:cot-sums}

\subsection{Asymptotics of the cotangent sum}
\label{appsec:asymptotics}

Here we derive the large-$N$ asymptotics of the bandwidth introduced in Sec.~\ref{sec:spectrum_structure}. Using the expression for $W$ given there, we need only evaluate
\begin{equation}
  S_N := \sum_{q=1}^{\lfloor \frac{N-1}2\rfloor} \cot\!\left(\frac{\pi q}{N}\right)
\end{equation}
for large $N$.

We decompose
\begin{equation}
  \cot x = \frac{1}{x} + \Bigl(\cot x - \frac{1}{x}\Bigr),
  \label{eq:cot_decomposition}
\end{equation}
and set $x_q=\pi q/N$. This yields
\begin{equation}
  S_N
  = \frac{N}{\pi}\sum_{q=1}^{M}\frac{1}{q}
    + \sum_{q=1}^{M}\Bigl(\cot x_q - \frac{1}{x_q}\Bigr),
  \label{eq:SN_split}
\end{equation}
where
\begin{equation}
  M\equiv\left\lfloor \frac{N-1}{2}\right\rfloor .
  \label{eq:M_def_asymp}
\end{equation}
The first term is expressed using the harmonic number
$H_M=\sum_{q=1}^M 1/q$:
\begin{equation}
  \frac{N}{\pi}\sum_{q=1}^{M}\frac{1}{q}
  =
  \frac{N}{\pi}H_M .
  \label{eq:harmonic_first_term}
\end{equation}
Using
\begin{equation}
  H_M=\ln M+\gamma+O(1/M),
  \label{eq:HM_asymp}
\end{equation}
where $\gamma$ is the Euler--Mascheroni constant, we obtain
\begin{equation}
  \frac{N}{\pi}H_M
  =
  \frac{N}{\pi}\bigl(\ln M+\gamma+O(1/M)\bigr).
  \label{eq:first_term_HM_asymp}
\end{equation}
Since $M\sim N/2$, this becomes
\begin{equation}
  \frac{N}{\pi}H_M
  =
  \frac{N}{\pi}\bigl(\ln N+\gamma-\ln 2\bigr)+O(1).
  \label{eq:first_term_largeN}
\end{equation}

The second term is regular at $x=0$ because $\cot x-1/x=O(x)$, so it can be approximated by a Riemann sum:
\begin{equation}
  \sum_{q=1}^{M}\Bigl(\cot x_q - \frac{1}{x_q}\Bigr)
  =
  \frac{N}{\pi}\int_{0}^{\pi/2}\!\Bigl(\cot x - \frac{1}{x}\Bigr)\,dx+O(1).
  \label{eq:second_term_Riemann}
\end{equation}
Using, for $\varepsilon>0$,
\begin{equation}
  \int_{\varepsilon}^{\pi/2}\cot x\,dx
  =
  \bigl[\ln\sin x\bigr]_{\varepsilon}^{\pi/2},
  \label{eq:int_cot_reg}
\end{equation}
and
\begin{equation}
  \int_{\varepsilon}^{\pi/2}\frac{dx}{x}
  =
  \bigl[\ln x\bigr]_{\varepsilon}^{\pi/2},
  \label{eq:int_invx_reg}
\end{equation}
and then taking the limit $\varepsilon\to0^+$, we find
\begin{equation}
  \int_{0}^{\pi/2}\!\Bigl(\cot x - \frac{1}{x}\Bigr)\,dx
  =
  \ln\frac{2}{\pi}.
  \label{eq:cot_minus_invx_integral}
\end{equation}

Therefore,
\begin{align}
  S_N
  &=\frac{N}{\pi}\bigl(\ln N+\gamma-\ln 2\bigr)
   +\frac{N}{\pi}\ln\frac{2}{\pi}+O(1)\nonumber\\
  &=\frac{N}{\pi}\bigl(\ln N+\gamma-\ln\pi\bigr)+O(1),
  \label{eq:SN_asymp}
\end{align}
and hence
\begin{equation}
  W = 2S_N
    =\frac{2N}{\pi}\bigl(\ln N+\gamma-\ln\pi\bigr)+O(1).
    \label{eq:W_asymp}
\end{equation}
This is the asymptotic form used in Sec.~\ref{sec:spectrum_structure}.

\subsection{Low-temperature expansion of the free energy}
\label{appsec:lowT_thermo}

Here we derive the low-temperature expansion quoted in Eq.~\eqref{eq:lowT}.
Throughout this subsection, we use the notation $M$ defined in Eq.~\eqref{eq:M_def_asymp}.

For simplicity, here we consider odd $N$.
Starting from the partition function in Eq.~\eqref{eq:Zbeta}, the active-sector contribution can be rewritten as
\begin{equation}
\prod_{q=1}^{M}4\cosh^2\!\left(\frac{\beta\epsilon_q}{2}\right)
=
\exp\!\left(\beta\sum_{q=1}^{M}\epsilon_q\right)
\prod_{q=1}^{M}\bigl(1+e^{-\beta\epsilon_q}\bigr)^2,
\label{eq:ZlowT_pair}
\end{equation}
where we used $4\cosh^2(x/2)=e^x(1+e^{-x})^2$.

The frozen-sector contribution  in Eq.~\eqref{eq:Zbeta} is
\begin{equation}
4^M=\exp \bigl(N\log 2+O(1)\bigr).
\label{eq:ZlowT_zero}
\end{equation}
By Eq.~\eqref{eq:SN_asymp} and $\epsilon_q=\cot(\theta_q/2)$, we have
\begin{equation}
\sum_{q=1}^{M}\epsilon_q
=
\frac{N}{\pi}\bigl(\log N+\gamma-\log\pi\bigr)+O(1).
\label{eq:sumepsilon_app}
\end{equation}
Thus, at fixed $\beta>0$ and large $N$, the active sector dominates over the frozen sector, and
\begin{equation}
\log Z(\beta)
=
\beta\sum_{q=1}^{M}\epsilon_q
+
2\sum_{q=1}^{M}\log \bigl(1+e^{-\beta\epsilon_q}\bigr)
+o(N).
\label{eq:logZ_lowT_app}
\end{equation}
Therefore,
\begin{equation}
\frac{F(T,N)}{N}
=
-\frac{1}{N}\sum_{q=1}^{M}\epsilon_q
-\frac{2}{\beta N}\sum_{q=1}^{M}\log \bigl(1+e^{-\beta\epsilon_q}\bigr)
+o(1).
\label{eq:F_lowT_split_app}
\end{equation}

The first term is fixed by Eq.~\eqref{eq:sumepsilon_app}. For the second term, the low-temperature contribution comes from modes near $\theta=\pi$, where $\epsilon_q=\cot(\theta_q/2)$ is small. Writing $q=(N-1)/2-m$, we obtain
\begin{equation}
\begin{aligned}
\epsilon_{(N-1)/2-m}
&=\cot\left[\frac{\pi}{N}\left(\frac{N-1}{2}-m\right)\right]
=\tan\left[\frac{\pi}{N}\left(m+\frac12\right)\right]\\
&=\frac{\pi}{N}\left(m+\frac12\right)
+O\left(\frac{(m+1/2)^3}{N^3}\right).   
\end{aligned}
\end{equation}
For large $\beta$, after taking $N\to\infty$, we find
\begin{align}
\lim_{N\to\infty}&\frac{1}{N}\sum_{q=1}^{M}\log\!\bigl(1+e^{-\beta\epsilon_q}\bigr)\nonumber\\
&\simeq
\lim_{N\to\infty}\frac{1}{N}\sum_{m=0}^{\infty}
\log\!\left(1+e^{-(\beta\pi/N)(m+1/2)}\right)
\nonumber\\
&=
\frac{1}{\pi}\int_0^\infty du\,\log(1+e^{-\beta u}),
\end{align}
which, after rescaling $v=\beta u$, gives
\begin{equation}
\frac{1}{\beta\pi}\int_0^\infty dv\,\log(1+e^{-v}).
\end{equation}
Using the standard integral %Appendix~\ref{appsec:FD_integral},
\begin{equation}
\label{eq:standard_integral}
\int_0^\infty dv\,\log(1+e^{-v})=\frac{\pi^2}{12},
\end{equation}
we obtain
\begin{equation}
\lim_{N\to\infty}\frac{1}{N}\sum_{q=1}^{M}\log\!\bigl(1+e^{-\beta\epsilon_q}\bigr)
=
\frac{\pi}{12}\frac{1}{\beta}+O(\beta^{-3}).
\label{eq:logsum_lowT_app}
\end{equation}
Substituting Eqs.~\eqref{eq:sumepsilon_app} and \eqref{eq:logsum_lowT_app} into Eq.~\eqref{eq:F_lowT_split_app}, we obtain
\begin{equation}
\frac{F(T,N)}{N}
=
-\frac{1}{\pi}\bigl(\log N+\gamma-\log\pi\bigr)
-\frac{\pi}{6}T^2
+O(T^4),
\end{equation}
and hence
\begin{equation}
\frac{S}{N}
=
-\frac{\partial}{\partial T}\frac{F}{N}
=
\frac{\pi}{3}T+O(T^3).
\end{equation}
This reproduces Eq.~\eqref{eq:lowT} and the entropy formula quoted in Sec.~\ref{sec:thermo}.

\subsection{Sum formula for products of $\cot^2$}
\label{appsec:sumcot}

In this subsection, we derive the identity
\begin{equation}
\sum_{\{ q_{i<} \}}
\prod_{m=1}^{j}\cot^2\!\left(\frac{\pi q_{i_m}}{N}\right)
=
\frac{1}{N}\binom{N}{2j+1},
\label{sumcot}
\end{equation}
where $\sum_{\{ q_{i<} \}}$ denotes the sum over all strictly ordered
$j$-tuples of integers satisfying
\begin{equation}
1\le q_{i_1}<q_{i_2}<\cdots<q_{i_j}\le M,
\quad
M=\left\lfloor \frac{N-1}{2}\right\rfloor .
\end{equation}

We first consider odd $N=2n+1$. Then $M=n$, and Eq.~\eqref{sumcot} becomes
\begin{equation}
\sum_{\{ q_{i<} \}}
\prod_{m=1}^{j}\cot^2\!\left(\frac{\pi q_{i_m}}{2n+1}\right)
=
\frac{1}{2n+1}\binom{2n+1}{2j+1}.
\label{sumcotodd}
\end{equation}
Here the summation runs over $1\le q_{i_1}<\cdots<q_{i_j}\le n$.

Let
\begin{equation}
\theta_q=\frac{\pi q}{2n+1},
\qquad q=1,\dots,n.
\end{equation}
From
\begin{equation}
e^{{\rm i}(2n+1)\theta_q}
=
(\cos\theta_q + {\rm i}\sin\theta_q)^{2n+1},
\end{equation}
the imaginary part gives
\begin{equation}
0=
\sum_{k=0}^{n}
(-1)^{\,n-k}
\binom{2n+1}{2k}
(\cos\theta_q)^{2k}(\sin\theta_q)^{2(n-k)+1}.
\end{equation}
Dividing by $\sin^{2n+1}\theta_q$, we obtain
\begin{equation}
0=
\sum_{k=0}^{n}
(-1)^{\,n-k}
\binom{2n+1}{2k}
(\cot\theta_q)^{2k}.
\end{equation}
Thus $(\cot\theta_q)^2$ are roots of the polynomial
\begin{equation}
f(x)=\sum_{k=0}^{n}(-1)^{\,n-k}\binom{2n+1}{2k}x^k.
\end{equation}
Since the coefficient of $x^n$ is $2n+1$, we can write
\begin{equation}
f(x)=(2n+1)\prod_{q=1}^{n}\bigl[x-(\cot\theta_q)^2\bigr].
\end{equation}
Comparing the coefficients of $x^{n-j}$ yields Eq.~\eqref{sumcotodd}.

For even $N=2(n+1)$, we have $M=n$, and Eq.~\eqref{sumcot} becomes
\begin{equation}
\sum_{\{ q_{i<} \}}
\prod_{m=1}^{j}\cot^2\!\left(\frac{\pi q_{i_m}}{2(n+1)}\right)
=
\frac{1}{2(n+1)}\binom{2(n+1)}{2j+1}.
\label{sumcoteven}
\end{equation}
Here the summation runs over $1\le q_{i_1}<\cdots<q_{i_j}\le n$.

Let
\begin{equation}
\theta_q=\frac{\pi q}{2(n+1)},
\qquad q=1,\dots,n.
\end{equation}
Taking the imaginary part of
\begin{equation}
e^{{\rm i}2(n+1)\theta_q}
=
(\cos\theta_q+{\rm i}\sin\theta_q)^{2(n+1)}
\end{equation}
gives
\begin{equation}
0=
\sum_{k=0}^{n}
(-1)^{\,n-k}
\binom{2(n+1)}{2k+1}
(\cot\theta_q)^{2k}.
\end{equation}
The same coefficient comparison yields Eq.~\eqref{sumcoteven}, and therefore
Eq.~\eqref{sumcot}. This completes the proof of Eq.~\eqref{sumcot}.
\begin{comment}
\subsection{Evaluation of a standard integral}
\label{appsec:FD_integral}

Here we evaluate
\begin{equation}
I=\int_0^\infty dv\,\log(1+e^{-v}),
\end{equation}
which appears in the low-temperature expansion. After an integration by parts,
\begin{equation}
I
=
\int_0^\infty dv\,\frac{v}{e^v+1}.
\end{equation}
Using
\begin{equation}
\frac{1}{e^v+1}
=
\sum_{n=1}^{\infty}(-1)^{n-1}e^{-nv},
\qquad (v>0),
\end{equation}
we obtain
\begin{equation}
I
=
\sum_{n=1}^{\infty}(-1)^{n-1}\int_0^\infty dv\,v e^{-nv}
=
\sum_{n=1}^{\infty}\frac{(-1)^{n-1}}{n^2}.
\end{equation}
Using the Riemann zeta function,
\begin{equation}
\zeta(s):=\sum_{n=1}^{\infty}\frac{1}{n^s},
\qquad \Re(s)>1,
\end{equation}
together with
\begin{equation}
\sum_{n=1}^{\infty}\frac{(-1)^{n-1}}{n^s}
=
(1-2^{1-s})\zeta(s),
\end{equation}
we find
\begin{equation}
I=\frac{1}{2}\zeta(2)=\frac{\pi^2}{12}.
\end{equation}
Therefore,
\begin{equation}
\int_0^\infty dv\,\log(1+e^{-v})=\frac{\pi^2}{12}.
\end{equation}
\end{comment}

\section{Technical details for the spectral form factor}
\label{app:tech}

\subsection{Small-$t$ expansion at infinite temperature}
\label{appsec:sff_smallt}

Here we derive the small-$t$ expansion of the infinite-temperature SFF quoted in Eq.~\eqref{eq:g_small_t_main}.
At $\beta=0$,
\begin{equation}
\mathrm{Tr}\,e^{{\rm i}Ht}
=
\mathrm{Tr}\,\mathbb{I}
-\frac{t^2}{2}\mathrm{Tr}\,H^2
+\frac{t^4}{24}\mathrm{Tr}\,H^4
+O(t^6),
\end{equation}
and all odd moments vanish because the spectrum is symmetric under $H\to -H$.

We present the derivation for odd $N$. For even $N$, the unpaired mode $q=N/2$ contributes only an overall factor of $2$, independent of $t$, which cancels in the normalized SFF.

Using
\begin{equation}
H=H_{\rm pair}n_0,
\qquad
n_0^2=n_0,
\qquad
[H_{\rm pair},n_0]=0,
\end{equation}
we have
\begin{equation}
H^m=(H_{\rm pair})^m n_0,
\qquad m\ge 1.
\end{equation}
Tracing over $n_0\in\{0,1\}$ then gives
\begin{equation}
\mathrm{Tr}\,H^m=\mathrm{Tr}_{\mathrm{pair}}\!\left(H_{\rm pair}^m\right),
\qquad m\ge 1,
\label{eq:Tr_factor_correct}
\end{equation}
where $\mathrm{Tr}_{\mathrm{pair}}$ denotes the trace over the paired Bogoliubov modes $\{n_{q\sigma}\}$ with $q=1,\dots,M$ and $\sigma=\pm1$. Meanwhile,
\begin{equation}
\mathrm{Tr}\,\mathbb{I}=2^N.
\end{equation}

Let $M=\left\lfloor\frac{N-1}{2}\right\rfloor$. Then
\begin{equation}
H_{\rm pair}
=
\sum_{q=1}^{M}\epsilon_q\,(n_{q+}-n_{q-}),
\qquad
\epsilon_q=\cot\!\left(\frac{\pi q}{N}\right).
\end{equation}
Since different $q$ pairs are independent, for fixed $q$ one has
\begin{equation}
\sum_{n_{q+},n_{q-}=0}^{1}(n_{q+}-n_{q-})^k
=
\begin{cases}
0, & k\ \mathrm{odd},\\
2, & k\ \mathrm{even},
\end{cases}
\qquad k\ge 1 .
\label{eq:pair_sums}
\end{equation}

For the second moment, all cross terms with $q\neq r$ vanish after tracing, so
\begin{align}
\mathrm{Tr}_{\mathrm{pair}}\!\left(H_{\rm pair}^2\right)
&=
\sum_{q=1}^{M}\epsilon_q^2
\Bigl(\sum_{n_{q+},n_{q-}}(n_{q+}-n_{q-})^2\Bigr)
\prod_{r\neq q}\Bigl(\sum_{n_{r+},n_{r-}}1\Bigr)
\nonumber\\
&=
\sum_{q=1}^{M}\epsilon_q^2\cdot 2 \cdot 4^{M-1}
=
2^{2M-1}\sum_{q=1}^{M}\epsilon_q^2 .
\label{eq:TrH2_pre}
\end{align}
Using Eq.~\eqref{sumcot} from Appendix~\ref{appsec:sumcot} with $j=1$, we obtain
\begin{equation}
\sum_{q=1}^{M}\epsilon_q^2
=
\sum_{q=1}^{M}\cot^2\!\left(\frac{\pi q}{N}\right)
=
\frac{1}{N}\binom{N}{3},
\end{equation}
and hence
\begin{equation}
\mathrm{Tr}_{\mathrm{pair}}\!\left(H_{\rm pair}^2\right)
=
2^{2M-1}\,\frac{1}{N}\binom{N}{3}.
\label{eq:TrH2}
\end{equation}

For the fourth moment, the surviving terms are those in which either a single mode appears to the fourth power or two distinct modes each appear to the second power. 
This gives
\begin{align}
\mathrm{Tr}_{\mathrm{pair}}\!\left(H_{\rm pair}^4\right)
&=
2^{2M-1}
\left[
\sum_{q=1}^{M}\epsilon_q^4
+
3\sum_{1\le q<r\le M}\epsilon_q^2\epsilon_r^2
\right]
\nonumber\\
&=
2^{2M-1}
\left[
\left(\sum_{q=1}^{M}\epsilon_q^2\right)^2
+
\sum_{1\le q<r\le M}\epsilon_q^2\epsilon_r^2
\right].
\label{eq:TrH4_pre}
\end{align}
Using Eq.~\eqref{sumcot} again, now with $j=2$, we obtain
\begin{equation}
\sum_{1\le q<r\le M}\epsilon_q^2\epsilon_r^2
=
\frac{1}{N}\binom{N}{5}.
\end{equation}
Together with $\sum_q\epsilon_q^2=\frac{1}{N}\binom{N}{3}$, this gives
\begin{equation}
\mathrm{Tr}_{\mathrm{pair}}\!\left(H_{\rm pair}^4\right)
=
2^{2M-1}
\left[
\frac{1}{N^2}\binom{N}{3}^2
+
\frac{1}{N}\binom{N}{5}
\right].
\label{eq:TrH4}
\end{equation}

Using
\begin{equation}
g(t,0)=\left|\frac{\mathrm{Tr}\,e^{{\rm i}Ht}}{2^N}\right|^2
\end{equation}
and substituting Eqs.~\eqref{eq:Tr_factor_correct}, \eqref{eq:TrH2}, and
\eqref{eq:TrH4}, we obtain
\begin{equation}
g(t,0)
=
1
-\frac{t^2}{4N}\binom{N}{3}
+\frac{t^4}{24}\,\frac{1}{4N}
\left[
\frac{7}{2N}\binom{N}{3}^2
+
2\binom{N}{5}
\right]
+O(t^6).
\label{eq:g_small_t_final}
\end{equation}
This reproduces Eq.~\eqref{eq:g_small_t_main}.

\subsection{Exact representation and the plateau}
\label{appsec:sff_plateau}

The plateau of the infinite-temperature SFF can be understood directly from the
exact partition function. For odd $N$, at $\beta=0$,
\begin{equation}
Z(-{\rm i}t)=\mathrm{Tr}\,e^{{\rm i}Ht}
=
4^{M}
+
\prod_{q=1}^{M}4\cos^2\!\left(\frac{\epsilon_q t}{2}\right),
\end{equation}
where $M=\left\lfloor\frac{N-1}{2}\right\rfloor$. Since $Z(0)=2^N=2\cdot4^M$, this can be written as
\begin{equation}
Z(-{\rm i}t)
=
\frac{Z(0)}{2}\left[1+P_N^2\!\left(\frac{t}{2}\right)\right],
\end{equation}
where
\begin{equation}
P_N(x)=\prod_{q=1}^{M}\cos(\epsilon_q x).
\label{eq:PN_def_app}
\end{equation}
Therefore
\begin{equation}
g(t,0)
=
\frac{1}{4}\left[1+P_N^2\!\left(\frac{t}{2}\right)\right]^2.
\label{eq:g_SFF_PN}
\end{equation}

This form makes the origin of the plateau transparent. The first term in $Z(-{\rm i}t)$ comes from the extensive frozen sector and is strictly time independent. The second term comes from the active sector and is controlled by $P_N(x)$. 
As shown in Appendix~\ref{appsec:PN}, $P_N(x)$ is exponentially suppressed at large $N$ for any fixed $x\neq 0$. It follows that for fixed $t>0$,
\begin{equation}
P_N\!\left(\frac{t}{2}\right)\to 0,
\qquad N\to\infty,
\end{equation}
and Eq.~\eqref{eq:g_SFF_PN} reduces to
\begin{equation}
g(t,0)\xrightarrow{N\to\infty}\frac{1}{4}.
\end{equation}
This is the plateau value %quoted 
presented in Sec.~\ref{sec:sff}.

\subsection{Large-$N$ estimate of $P_N(x)$}
\label{appsec:PN}

Here we estimate the large-$N$ behavior of $P_N(x)$ defined in
Eq.~\eqref{eq:PN_def_app}. Following Ref.~\cite{ozaki2025disorder}, we introduce
the single-particle density of states
\begin{align}
D_1(\epsilon)
&=
\frac{1}{2M+1}\sum_{q=1}^{M}
\Bigl[\delta(\epsilon-\epsilon_q)+\delta(\epsilon+\epsilon_q)\Bigr]
\nonumber\\
&\simeq
\int_{-1/2}^{1/2}dq\,
\delta\!\left(\epsilon-\cot(\pi q)\right)
=
\frac{1}{\pi}\frac{1}{1+\epsilon^2},
\end{align}
where $M=\lfloor (N-1)/2\rfloor$.

Replacing the sum over $q$ by an integral, we obtain
\begin{equation}
\log|P_N(x)|
\simeq
\frac{N}{2\pi}\int_{-\infty}^{\infty}d\epsilon\,
\frac{1}{1+\epsilon^2}\log|\cos(\epsilon x)|.
\end{equation}
Evaluating the integral gives
\begin{equation}
|P_N(x)|
\simeq
\left(\frac{1+e^{-2|x|}}{2}\right)^{N/2},
\label{eq:PN-closed}
\end{equation}
Thus, for any fixed $x\neq 0$, $\left|P_N(x)\right|$ is exponentially suppressed as $N$ increases, explaining the rapid approach to the plateau in Eq.~\eqref{eq:g_SFF_PN}.

\section{Details of dynamics and correlation functions}
\label{app:dyn-details}

In this appendix, we derive the projector representation of time evolution, the sector decomposition of the thermal state and its weights, the momentum-space Green functions, and the complex-time representation and Wick reduction of the regulated OTOC.

\subsection{Projector representation of time evolution}
\label{app:U-sector}

Here we derive the projector representation of time evolution used in Sec.~\ref{sec:dynamics}. From Eq.~\eqref{eq:Hre} and the definitions in Eq.~\eqref{eq:P01_main}, we have
\begin{equation}
HP_0=0,\qquad HP_1=H_{\rm pair}P_1,
\end{equation}
which immediately gives
\begin{equation}
e^{-{\rm i}Ht}=P_0+P_1e^{-{\rm i}H_{\rm pair}t}.
\label{eq:U-sector}
\end{equation}
Accordingly, for any operator $\mathcal O$,
\begin{align}
\mathcal O(t)
&=e^{{\rm i}Ht}\mathcal O e^{-{\rm i}Ht}
\nonumber\\
&=
P_0\,\mathcal O\,P_0
+ e^{{\rm i}H_{\rm pair}t}P_1\,\mathcal O\,P_1 e^{-{\rm i}H_{\rm pair}t}
\nonumber\\
&\quad+
P_0\,\mathcal O\,P_1 e^{-{\rm i}H_{\rm pair}t}
+
e^{{\rm i}H_{\rm pair}t}P_1\,\mathcal O\,P_0.
\label{eq:Oevo_projector}
\end{align}
This yields the sector decomposition of operator evolution discussed in Sec.~\ref{sec:dynamics}.
\subsection{Partition function and sector decomposition}
\label{app:rho-sector}
This subsection derives the sector weights in Eq.~\eqref{eq:sector_weight_def} and the explicit expression for $w_1(\beta,N)$ used in Sec.~\ref{sec:dynamics}. Using Eqs.~\eqref{eq:Hre} and \eqref{eq:P01_main}, we obtain
\begin{equation}
e^{-\beta H}=P_0+P_1e^{-\beta H_{\rm pair}}.
\label{eq:boltz-sector}
\end{equation}
Therefore,
\begin{equation}
Z=\mathrm{Tr}\,e^{-\beta H}
=\mathrm{Tr}(P_0)+\mathrm{Tr}\!\left(P_1e^{-\beta H_{\rm pair}}\right),
\label{eq:Zsplit_app}
\end{equation}
Equation~\eqref{eq:boltz-sector} therefore implies
\begin{equation}
\rho(\beta)=\frac{e^{-\beta H}}{Z(\beta)}=w_0(\beta,N)\,\rho^{(0)}+w_1(\beta,N)\,\rho^{(1)},
\label{eq:rho_mix_app}
\end{equation}
with
\begin{equation}
\rho^{(0)}=\frac{P_0}{\mathrm{Tr}(P_0)},
\qquad
\rho^{(1)}=\frac{P_1e^{-\beta H_{\rm pair}}}{\mathrm{Tr}\!\left(P_1e^{-\beta H_{\rm pair}}\right)}.
\label{eq:rho_sector_app}
\end{equation}
For even $N$, the unpaired mode at $q=N/2$ contributes an additional factor of $2$, independent of $\beta$, which cancels in normalized quantities.

\subsection{Momentum-space correlators and complex-time extension}
\label{app:2pt-momentum}
Here we derive the momentum-space correlators quoted in Eqs.~\eqref{eq:ffnorm} and \eqref{eq:ffanom}.
Using Eq.~\eqref{eq:bogoliubov}, the Bogoliubov quasiparticles evolve as
\begin{equation}
\gamma_{q\sigma}(t)=e^{-{\rm i}\sigma\epsilon_q t}\gamma_{q\sigma},
\end{equation}
and their occupations in the state $\rho^{(1)}$ are
\begin{equation}
\langle n_{q\sigma}\rangle^{1}
=
\frac{1}{1+e^{\beta\sigma\epsilon_q}}
=
\frac12\left(1-\sigma\tanh\frac{\beta\epsilon_q}{2}\right).
\label{eq:nqsigma_app}
\end{equation}
Using
\begin{equation}
f_q=\frac{1}{\sqrt2}\left(\gamma_{q+}+\gamma_{q-}\right),
\qquad
f_{N-q}^\dagger=\frac{\rm i}{\sqrt2}\left(\gamma_{q+}-\gamma_{q-}\right),
\end{equation}
we obtain
\begin{align}
\big\langle f_q^\dagger(t)f_q(0)\big\rangle^{1}
&=
\frac12\Bigl[\cos(\epsilon_q t)
-{\rm i}\tanh\Bigl(\frac{\beta\epsilon_q}{2}\Bigr)\sin(\epsilon_q t)\Bigr],
\\
\big\langle f_q(t)f_{N-q}(0)\big\rangle^{1}
&=
-\frac12\Bigl[\sin(\epsilon_q t)
+{\rm i}\tanh\Bigl(\frac{\beta\epsilon_q}{2}\Bigr)\cos(\epsilon_q t)\Bigr],
\end{align}
which reproduce Eqs.~\eqref{eq:ffnorm} and \eqref{eq:ffanom} in the main text.
The remaining components follow from Hermitian conjugation and fermionic antisymmetry.

Since these two correlators are finite sums of $\sin(\epsilon_q t)$ and $\cos(\epsilon_q t)$, they extend directly to complex time $t\to t+{\rm i}\tau$, which is the extension used in the regulated OTOC.

For the frozen sector, the nonzero-momentum contractions are
\begin{equation}
\langle f_p^\dagger f_{p'}\rangle^{0}
=
\frac12\,\delta_{pp'},
\qquad
\langle f_p f_{p'}\rangle^{0}=0,
\qquad
(p,p'\neq 0).
\label{eq:sector0_app}
\end{equation}

\subsection{Real-space correlators at finite $N$}
\label{app:2pt-realspace}

Using the inverse Fourier transform, we obtain the real-space correlators from the momentum-space expressions in Appendix~\ref{app:2pt-momentum}.

We begin with the anomalous correlator $\langle c_j(t)c_k(0)\rangle$. Substituting the Fourier expansion, we have
\begin{equation}
\langle c_j(t)c_k(0)\rangle
=
\frac{1}{N}\sum_{p,q}
e^{{\rm i}\theta_p j}e^{{\rm i}\theta_q k}\,
\langle f_p(t)f_q(0)\rangle .
\end{equation}
Since $\langle f_p(t)f_q(0)\rangle$ is nonzero only when $q=N-p$, we obtain
\begin{align}
\langle c_j(t)c_k(0)\rangle
=
\frac{{\rm i}\,w_1}{N}&\sum_{q=1}^{\lfloor (N-1)/2\rfloor}
\sin\!\big[\theta_q(j-k)\big]\nonumber\\
&\times\Bigl[
-\sin(\epsilon_q t)
-{\rm i}\tanh \Bigl(\frac{\beta\epsilon_q}{2}\Bigr)\cos(\epsilon_q t)
\Bigr].
\label{eq:Ccc_app}
\end{align}
Note that the $q=0$ mode does not contribute. For even $N$, the unpaired mode $q=N/2$ also drops out because its anomalous correlator vanishes.
It follows that
\begin{equation}
\langle c_j^\dagger(t)c_k^\dagger(0)\rangle
=
-\langle c_j(t)c_k(0)\rangle .
\label{eq:Cdagdag_app}
\end{equation}

We next consider the mixed correlators. In this case, the $q=0$ mode contributes explicitly, and for even $N$ the unpaired mode $q=N/2$ gives an additional term of order $1/N$.
One finds
\begin{align}
\langle c_j(t)c_k^\dagger(0)\rangle
&=
\frac{w_0}{2}\,\delta_{jk}+\frac{w_0}{N}\!\left[
P_N^2\!\left(\frac{t}{2}\right)-\frac12
\right]
\nonumber\\
&\quad+\frac{w_1}{N}\sum_{q=1}^{\lfloor (N-1)/2\rfloor}
\cos\!\big[\theta_q(j-k)\big]\nonumber\\
&\qquad\qquad \times\Bigl[
\cos(\epsilon_q t)
-{\rm i}\tanh\!\Bigl(\frac{\beta\epsilon_q}{2}\Bigr)\sin(\epsilon_q t)
\Bigr]
\nonumber\\
&\quad
+ %\delta_{N,\mathrm{even}}\,
\frac{1+(-1)^N}{2}\,
\frac{w_1}{2N}e^{{\rm i}\pi(j-k)},
\label{eq:CcCdag_app}
\end{align}
and
\begin{align}
\langle c_j^\dagger(t)c_k(0)\rangle
&=
\frac{w_0}{2}\,\delta_{jk}-\frac{w_0}{2N}
\nonumber\\
&\quad+\frac{w_1}{N}\sum_{q=1}^{\lfloor (N-1)/2\rfloor}
\cos\!\big[\theta_q(j-k)\big]\nonumber\\
&\qquad\qquad \times\Bigl[
\cos(\epsilon_q t)
-{\rm i}\tanh\Bigl(\frac{\beta\epsilon_q}{2}\Bigr)\sin(\epsilon_q t)
\Bigr]
\nonumber\\
&\quad
+\frac{w_1}{N}
\frac{P_N^2\!\bigl(\frac{t+{\rm i}\beta}{2}\bigr)}
     {P_N^2\!\bigl(\frac{{\rm i}\beta}{2}\bigr)}
+%\delta_{N,\mathrm{even}}\,
\frac{1+(-1)^N}{2}\,
\frac{w_1}{2N}e^{{\rm i}\pi(j-k)} .
\label{eq:CdagC_app}
\end{align}
Here $P_N(x)$ is defined in Eq.~\eqref{eq:PN_def_app}. In both equations, the last term appears only for even $N$ and comes from the unpaired mode $q=N/2$.

\subsection{Evaluation of the regulated OTOC}
\label{app:otoc-evaluation}

We now evaluate the component expansion in
Eq.~\eqref{eq:OTOC_component_sum}. The operators defined in
Eq.~\eqref{eq:d_z0} satisfy
\begin{align}
P_\alpha d_jP_\gamma
&=\delta_{\alpha\gamma}P_\alpha d_j,
&
P_\alpha d_j^\dagger P_\gamma
&=\delta_{\alpha\gamma}P_\alpha d_j^\dagger,
\nonumber\\
z_0&=P_0z_0P_1,
&
z_0^\dagger&=P_1z_0^\dagger P_0.
\label{eq:projector_action_dz_app}
\end{align}
For an ordered product
$\mathcal O_1\mathcal O_2\mathcal O_3\mathcal O_4$, we call
$(\alpha_0,\alpha_1,\alpha_2,\alpha_3)$ a nonzero projector path if
\begin{equation}
P_{\alpha_0}\mathcal O_1P_{\alpha_1}
\mathcal O_2P_{\alpha_2}
\mathcal O_3P_{\alpha_3}
\mathcal O_4P_{\alpha_0}
\neq0.
\label{eq:projector_path_def}
\end{equation}
Using Eq.~\eqref{eq:projector_action_dz_app}, the possible paths are
\begin{equation}
\begin{array}{c|c}
dddd &(0,0,0,0),\ (1,1,1,1)\\
zddz &(1,0,0,0)\\
zdzd &(1,0,0,1)\\
dzdz &(1,1,0,0)\\
dzzd &(1,1,0,1).
\end{array}
\label{eq:OTOC_paths_app}
\end{equation}
All other components vanish in the trace over the zero-momentum mode.
The $zddz$ and $dzzd$ contributions also vanish after the sums over
$j$ and $k$, because
\begin{equation}
\sum_{j=0}^{N-1}d_j
=
\sum_{j=0}^{N-1}d_j^\dagger
=0.
\label{eq:sum_d_app}
\end{equation}
This leaves the three contributions displayed in
Eq.~\eqref{eq:OTOC_surviving_terms}.

For the $dddd$ contribution, the two paths in
Eq.~\eqref{eq:OTOC_paths_app} give
\begin{equation}
\otoc_{dddd}(t;\beta)
=
w_0\,\otoc_{dddd}^{0}(t;\beta)
+
w_1\,\otoc_{dddd}^{1}(t;\beta),
\label{eq:OTOC_dddd_decomp_app}
\end{equation}
where
\begin{equation}
\otoc_{dddd}^{\alpha}(t;\beta)
=
\frac{1}{N^2}
\sum_{j,k=0}^{N-1}
\left\langle
d_j^\dagger(t_1)d_k^\dagger(t_2)
d_j(t_3)d_k(t_4)
\right\rangle^\alpha .
\label{eq:OTOC_dddd_sector_app}
\end{equation}

For later use, we define the two-point functions in each sector as
\begin{equation}
\left(G_{\mathcal X\mathcal Y}^{\alpha}\right)_{jk}(\tau;\beta')
\equiv
\frac{
\mathrm{Tr}\!\left[
P_\alpha e^{-\beta'H}
\mathcal X_j(\tau)\mathcal Y_k(0)
\right]}
{\mathrm{Tr}\!\left(P_\alpha e^{-\beta'H}\right)},
\qquad
\mathcal X,\mathcal Y\in\{d,d^\dagger\}.
\label{eq:Gd_sector_def}
\end{equation}
We keep $\beta'$ explicit, since both $\beta'=\beta$ and $\beta'=\beta/2$ occur below.

Using Wick's theorem in each sector, we obtain
\begin{align}
\otoc_{dddd}^{\alpha}(t;\beta)
&=
\frac{1}{N^2}
\sum_{j,k=0}^{N-1}
\Bigl[
\left(G_{d^\dagger d^\dagger}^{\alpha}\right)_{jk}(t_{12};\beta)
\left(G_{dd}^{\alpha}\right)_{jk}(t_{34};\beta)
\nonumber\\
&\qquad
-
\left(G_{d^\dagger d}^{\alpha}\right)_{jj}(t_{13};\beta)
\left(G_{d^\dagger d}^{\alpha}\right)_{kk}(t_{24};\beta)
\nonumber\\
&\qquad
+
\left(G_{d^\dagger d}^{\alpha}\right)_{jk}(t_{14};\beta)
\left(G_{d^\dagger d}^{\alpha}\right)_{kj}(t_{23};\beta)
\Bigr],
\label{eq:OTOC_dddd_app}
\end{align}
where $t_{ab}=t_a-t_b$ and the times $t_a$ are given in Eq.~\eqref{eq:otoc_times}.

The active-sector two-point functions follow by transforming the momentum-space correlators in Eqs.~\eqref{eq:ffnorm} and \eqref{eq:ffanom} to the $d_j$ basis, which excludes the zero-momentum mode. Replacing $\beta$ in those equations by $\beta'$, we obtain
\begin{align}
\left(G_{dd}^{1}\right)_{jk}(\tau;\beta')
&=
\frac{{\rm i}}{N}
\sum_{q=1}^{\lfloor(N-1)/2\rfloor}
\sin\bigl[\theta_q(j-k)\bigr]\nonumber\\
&\qquad\times
\Bigl[
-\sin(\epsilon_q\tau)
-{\rm i}\tanh\left(\frac{\beta'\epsilon_q}{2}\right)
\cos(\epsilon_q\tau)
\Bigr],
\label{eq:Gdd_active_app}\\
\left(G_{d^\dagger d}^{1}\right)_{jk}(\tau;\beta')
&=
\frac{1}{N}
\sum_{q=1}^{\lfloor(N-1)/2\rfloor}
\cos\bigl[\theta_q(j-k)\bigr]\nonumber\\
&\qquad\times
\Bigl[
\cos(\epsilon_q\tau)
-{\rm i}\tanh\left(\frac{\beta'\epsilon_q}{2}\right)
\sin(\epsilon_q\tau)
\Bigr]\nonumber\\
&\quad+
\frac{1+(-1)^N}{2}
\frac{e^{{\rm i}\pi(j-k)}}{2N}.
\label{eq:Gdagd_active_app}
\end{align}
The remaining two-point functions satisfy
\begin{equation}
\begin{aligned}
\left(G_{d^\dagger d^\dagger}^{1}\right)_{jk}(\tau;\beta')
=
-\left(G_{dd}^{1}\right)_{jk}(\tau;\beta'),
\\
\left(G_{dd^\dagger}^{1}\right)_{jk}(\tau;\beta')
=
\left(G_{d^\dagger d}^{1}\right)_{jk}(\tau;\beta').
\end{aligned}
\label{eq:G_active_relations_app}
\end{equation}

The frozen-sector contractions follow from Eq.~\eqref{eq:sector0_app}:
\begin{align}
\left(G_{d^\dagger d}^{0}\right)_{jk}(\tau;\beta')
&=
\left(G_{dd^\dagger}^{0}\right)_{jk}(\tau;\beta')
=
\frac12\left(\delta_{jk}-\frac1N\right),
\nonumber\\
\left(G_{dd}^{0}\right)_{jk}(\tau;\beta')
&=
\left(G_{d^\dagger d^\dagger}^{0}\right)_{jk}(\tau;\beta')
=0.
\label{eq:G_frozen_app}
\end{align}

We next evaluate $\otoc_{zdzd}$ and $\otoc_{dzdz}$ using the projector paths in Eq.~\eqref{eq:OTOC_paths_app}. For either path, the zero-momentum trace gives a factor $1/N$. The two remaining factors $e^{-\beta H_{\rm pair}/4}$ combine into $e^{-\beta H_{\rm pair}/2}$. The remaining trace is therefore an active-sector two-point function at inverse temperature $\beta/2$. We obtain 
\begin{equation}
\begin{aligned}
\otoc_{zdzd}(t;\beta)
&=
\otoc_{dzdz}(t;\beta)\\
&=
-\frac{Z_{\rm pair}(\beta/2)}
{N\left[4^{\lfloor(N-1)/2\rfloor}+Z_{\rm pair}(\beta)\right]}
\left(G_{d^\dagger d}^{1}\right)_{\ell\ell}
\left(
t-\frac{{\rm i}\beta}{4};
\frac{\beta}{2}
\right),  
\end{aligned}
\label{eq:OTOC_mixed_app}
\end{equation}
where the right-hand side is independent of $\ell$. Substituting Eq.~\eqref{eq:Gdagd_active_app} gives
\begin{eqnarray}
\left(G_{d^\dagger d}^{1}\right)_{\ell\ell}
\left(
t-\frac{{\rm i}\beta}{4};
\frac{\beta}{2}
\right)
&=&
\frac{1}{N}
\sum_{q=1}^{\lfloor(N-1)/2\rfloor}
\frac{\cos(\epsilon_qt)}
{\cosh(\beta\epsilon_q/4)}
\nonumber \\
&&+
\frac{1+(-1)^N}{2}\,
\frac{1}{2N}.
\label{eq:G_mixed_sum_app}
\end{eqnarray}

Combining Eqs.~\eqref{eq:OTOC_dddd_decomp_app}, \eqref{eq:OTOC_mixed_app}, and \eqref{eq:G_mixed_sum_app}, we obtain Eq.~\eqref{eq:OTOC_finiteN}.

At infinite temperature, $w_0=w_1=1/2$. The Wick contractions in Eq.~\eqref{eq:OTOC_dddd_app} give
\begin{equation}
\otoc_{dddd}^{0}(t;0)
=
\otoc_{dddd}^{1}(t;0)
=
-\frac{(N-1)(N-2)}{4N^2}.
\label{eq:OTOC_dddd_beta0_app}
\end{equation}
In the active sector, the possible time dependence cancels through $\sin^2(\epsilon_qt)+\cos^2(\epsilon_qt)=1$. The exact infinite-temperature result at finite $N$ is therefore
\begin{equation}
\otoc(t;0)
=
-\frac{N^2-3N+3+(-1)^N}{4N^2}
-
\frac{1}{N^2}
\sum_{q=1}^{\lfloor(N-1)/2\rfloor}
\cos(\epsilon_qt).
\label{eq:OTOC_beta0_finiteN_app}
\end{equation}
At $t=0$, this reduces to
$\otoc(0;0)=-(N-1)/(4N)$, in agreement with a direct evaluation using
the fermionic anticommutation relations.
\subsection{Large-$N$ limit of the two-point correlators and the regulated OTOC}
\label{app:dyn-largeN}

We now take the large-$N$ limit of the correlators in the original $c_j$ basis obtained in Appendix~\ref{app:2pt-realspace}. The contributions from the $q=0$ mode and, for even $N$, from the unpaired mode $q=N/2$ are all of order $1/N$, and therefore vanish as $N\to\infty$. By contrast, the sums over the paired modes remain finite and are replaced by integrals over $\theta\in[0,\pi]$, with $\epsilon(\theta)=\cot\!\left(\frac{\theta}{2}\right)$.
Applying this to Eq.~\eqref{eq:Ccc_app}, we obtain
\begin{align}
\langle c_j(t)c_k(0)\rangle
&\xrightarrow[N\to\infty]{}
{\rm i}\,\frac{w_1}{2\pi}\int_0^\pi d\theta\,
\sin\!\big[\theta(j-k)\big]\nonumber\\
&\qquad \times
\Bigl[
-\sin\!\bigl(\epsilon(\theta)t\bigr)
-{\rm i}\tanh \Bigl(\frac{\beta\epsilon(\theta)}{2}\Bigr)
\cos\!\bigl(\epsilon(\theta)t\bigr)
\Bigr].
\label{eq:Ccc_largeN_app}
\end{align}
Equation~\eqref{eq:Cdagdag_app} continues to hold in the large-$N$ limit.

Similarly, Eqs.~\eqref{eq:CcCdag_app} and \eqref{eq:CdagC_app} reduce to
\begin{align}
\langle c_j(t)c_k^\dagger(0)\rangle
&=
\langle c_j^\dagger(t)c_k(0)\rangle + O(N^{-1})
\nonumber\\
&\xrightarrow[N\to\infty]{}
\frac{w_0}{2}\,\delta_{jk}
+\frac{w_1}{2\pi}\int_0^\pi d\theta\,
\cos\!\big[\theta(j-k)\big]\nonumber\\
&\qquad \times
\Bigl[
\cos\!\bigl(\epsilon(\theta)t\bigr)
-{\rm i}\tanh \Bigl(\frac{\beta\epsilon(\theta)}{2}\Bigr)
\sin\!\bigl(\epsilon(\theta)t\bigr)
\Bigr].
\label{eq:Cmix_largeN_app}
\end{align}
Thus, although the two mixed correlators differ at finite $N$, they become equal
in the large-$N$ limit.

We next consider the OTOC. In Eq.~\eqref{eq:OTOC_dddd_app}, the sums over $j$ and $k$ in the first and third terms on the right-hand side each reduce to a single sum over $q$. Both terms are therefore of order $1/N$. The second term, by contrast, can contribute at $O(1)$. Since $t_{13}=t_{24}=-{\rm i}\beta/2$, this term is independent of $t$. Using Eqs.~\eqref{eq:Gdagd_active_app} and \eqref{eq:G_frozen_app}, we obtain
\begin{align}
\left(G_{d^\dagger d}^{0}\right)_{\ell\ell}
\left(-\frac{{\rm i}\beta}{2};\beta\right)
&\xrightarrow[N\to\infty]{}\frac12,
\nonumber\\
\left(G_{d^\dagger d}^{1}\right)_{\ell\ell}
\left(-\frac{{\rm i}\beta}{2};\beta\right)
&\xrightarrow[N\to\infty]{}
\frac{1}{\pi}\int_0^\infty
\frac{d\epsilon}
{(1+\epsilon^2)\cosh(\beta\epsilon/2)}.
\label{eq:Gdagd_imaginary_largeN_app}
\end{align}
The two contributions in Eq.~\eqref{eq:OTOC_mixed_app} are of order $1/N$. It follows that
\begin{equation}
\otoc(t;\beta)
=
-\frac{w_0}{4}
-
\frac{w_1}{\pi^2}
\left[
\int_0^\infty
\frac{d\epsilon}
{(1+\epsilon^2)\cosh(\beta\epsilon/2)}
\right]^2
+
O\left(\frac{1}{N}\right),
\label{eq:OTOC_leading_largeN_app}
\end{equation}
where the displayed $O(1)$ contribution is independent of $t$.

At infinite temperature, the integral in Eq.~\eqref{eq:OTOC_leading_largeN_app} equals $\pi/2$. Since $w_0+w_1=1$, the leading contribution is $-1/4$. The time-dependent term can be obtained from Eq.~\eqref{eq:OTOC_beta0_finiteN_app}, using
\begin{align}
\frac{1}{N}\sum_{q=1}^{\lfloor(N-1)/2\rfloor}\cos(\epsilon_qt)
&\xrightarrow[N\to\infty]{}
\frac{1}{2\pi}\int_0^\pi d\theta\,
\cos\left[t\cot\left(\frac{\theta}{2}\right)\right]
\nonumber\\
&=
\frac{1}{\pi}\int_0^\infty d\epsilon\,
\frac{\cos(\epsilon t)}{1+\epsilon^2}
=
\frac12e^{-|t|}.
\label{eq:OTOC_mode_sum_largeN_app}
\end{align}
Together with
\begin{equation}
-\frac{(N-1)(N-2)}{4N^2}
=
-\frac14+\frac{3}{4N}
+
O\left(\frac{1}{N^2}\right),
\end{equation}
this gives Eq.~\eqref{eq:OTOC_beta0_largeN}. 

\section{Disorder-free one-dimensional breakdown interaction}
\label{app:gate-to-1d}

In this appendix, we show that a similar sector structure also appears in a disorder-free version of the one-dimensional quantum breakdown model without the chemical potential term. For uniform couplings, the same on-site Fourier transform used in the main text isolates a distinguished symmetric mode on each site, which provides a simple interpretation of the resulting Hilbert-space fragmentation.

We start from the original breakdown interaction~\cite{lian2023quantum},
\begin{equation}
  H_{\mathrm{I}}
  =
  \sum_{m=1}^{M-1}
  \sum_{1 \le i<j<k \le N}
  \sum_{l=1}^{N}
  \left(
    J^{ijk}_{m,l}\,
    c^{\dagger}_{m+1,i}
    c^{\dagger}_{m+1,j}
    c^{\dagger}_{m+1,k}
    c_{m,l}
    + \text{H.c.}
  \right),
  \label{eq:Lian-HI}
\end{equation}
where $m=1,\dots, M$ labels the site along the chain, $i,j,k,l=1,\dots,N$ label the internal fermion modes within each site,  and $J^{ijk}_{m,l}$ are complex parameters. 
We denote by $c_{m,i}$ and $c^\dagger_{m,i}$ the annihilation and creation operators of the $i$th fermion mode on the $m$th site. 

To define a disorder-free version, we choose a uniform coupling, $J^{ijk}_{m,l}\equiv J\in\mathbb R$. 
Then Eq.~\eqref{eq:Lian-HI} becomes
\begin{equation}
  H_{\mathrm{I}}^{(\mathrm{df})}
  =
  J\sum_{m=1}^{M-1}
  \sum_{1\le i<j<k\le N}
  \sum_{l=1}^{N}
  \Bigl(
    c^{\dagger}_{m+1,i}c^{\dagger}_{m+1,j}c^{\dagger}_{m+1,k}\,c_{m,l}
    + \text{H.c.}
  \Bigr).
  \label{eq:HI-df-raw}
\end{equation}

Next we perform the on-site Fourier transform:
\begin{equation}
  f_{p}^{(m)}
  \equiv \frac{1}{\sqrt{N}}\sum_{\ell=1}^{N} c_{m,\ell}\,
  e^{-{\rm i}\frac{2\pi}{N}p\ell},
  \qquad p=0,1,\dots,N-1.
  \label{eq:FT-site}
\end{equation}
Introducing the site-resolved analogue of Eq.~\eqref{eq:Q},
\begin{equation}
  A^{(m)}
  \equiv
  \sum_{1\le j<k\le N}c_{m,j}c_{m,k}
  =
  \frac12\sum_{j,k=1}^{N}\mathscr A_{jk}\,c_{m,j}c_{m,k},
  \label{eq:Am-def}
\end{equation}
we obtain
\begin{equation}
  H_{\mathrm{I}}^{(\mathrm{df})}
  =
  -N J\sum_{m=1}^{M-1}\Bigl(
  f_{0}^{\dagger(m+1)}\, f_{0}^{(m)}\, A^{\dagger(m+1)}
  + A^{(m+1)}\, f_{0}^{\dagger(m)}\, f_{0}^{(m+1)}
  \Bigr),
  \label{eq:HI-f0m-gated}
\end{equation}
which has the same basic structure as the factorized form in Eq.~\eqref{eq:Hre}, except that the distinguished modes now live on neighboring sites.

\begin{figure}[t]
  \centering
  \includegraphics[width=0.9\linewidth]{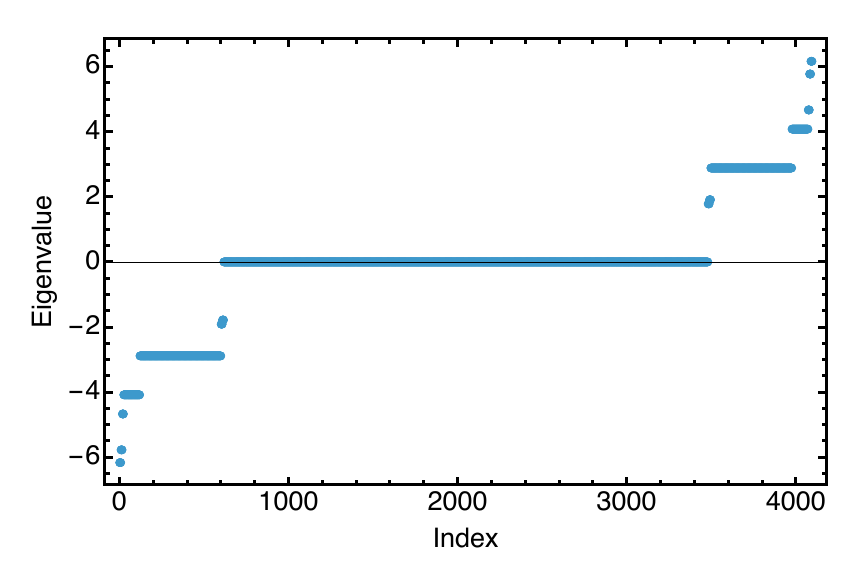}
  \caption{%Sorted 
  Eigenvalues of the disorder-free one-dimensional breakdown interaction
  $H_{\mathrm I}^{(\mathrm{df})}/N$, sorted in ascending order. 
  The system has $M=4$ sites and $N=3$ modes per site, with coupling $J=5$.}
  \label{fig:spectrum_HI_df}
\end{figure}

Defining the site-resolved occupation
\begin{equation}
  n_0^{(m)}\equiv f_0^{\dagger(m)}f_0^{(m)},
\end{equation}
we see from Eq.~\eqref{eq:HI-f0m-gated} that any intersite breakdown process necessarily involves the symmetric mode on the corresponding site. In particular, if $n_0^{(m)}=0$ on a given many-body basis state, then the forward process from site $m$ to $m+1$ is blocked, because the annihilation operator $f_0^{(m)}$ acts trivially on that state. 
As an extreme example, the subspace with $n_0^{(m)}=0$ for all $m$ is annihilated by $H_{\mathrm{I}}^{(\mathrm{df})}$, giving a large family of exact eigenstates and hence extensive degeneracies. More generally, the dynamics generated by $H_{\mathrm{I}}^{(\mathrm{df})}$ is strongly constrained by the pattern $\{n_0^{(m)}\}$, which provides a natural origin of the fragmentation structure in this disorder-free one-dimensional model; see Fig.~\ref{fig:spectrum_HI_df}.

Finally, in the original $N=3$ model, the special mode $f_{m,1}$ is defined by an on-site $U(3)$ rotation~\cite{lian2023quantum}. In the uniform-coupling disorder-free case, symmetry allows one to choose $f_{m,1}$ as the symmetric combination, that is, $f_{m,1}\equiv f_0^{(m)}$ up to a phase, whereas this identification does not hold for generic couplings.

\bibliography{refs}

\end{document}